\documentclass[article,nojss,shortnames]{jss}\usepackage[]{graphicx}\usepackage[]{xcolor}
\makeatletter
\def\maxwidth{ %
  \ifdim\Gin@nat@width>\linewidth
    \linewidth
  \else
    \Gin@nat@width
  \fi
}
\makeatother

\definecolor{fgcolor}{rgb}{0, 0, 0}
\newcommand{\hlnum}[1]{\textcolor[rgb]{0,0,0}{#1}}%
\newcommand{\hlsng}[1]{\textcolor[rgb]{0,0,0}{#1}}%
\newcommand{\hlcom}[1]{\textcolor[rgb]{0.4,0.4,0.4}{\textit{#1}}}%
\newcommand{\hlopt}[1]{\textcolor[rgb]{0,0,0}{\textbf{#1}}}%
\newcommand{\hldef}[1]{\textcolor[rgb]{0,0,0}{#1}}%
\newcommand{\hlkwa}[1]{\textcolor[rgb]{0,0,0}{\textbf{#1}}}%
\newcommand{\hlkwb}[1]{\textcolor[rgb]{0,0,0}{\textbf{#1}}}%
\newcommand{\hlkwc}[1]{\textcolor[rgb]{0,0,0}{\textbf{#1}}}%
\newcommand{\hlkwd}[1]{\textcolor[rgb]{0,0,0}{\textbf{#1}}}%

\usepackage{framed}
\makeatletter
\newenvironment{kframe}{%
 \def\at@end@of@kframe{}%
 \ifinner\ifhmode%
  \def\at@end@of@kframe{\end{minipage}}%
  \begin{minipage}{\columnwidth}%
 \fi\fi%
 \def\FrameCommand##1{\hskip\@totalleftmargin \hskip-\fboxsep
 \colorbox{shadecolor}{##1}\hskip-\fboxsep
     \hskip-\linewidth \hskip-\@totalleftmargin \hskip\columnwidth}%
 \MakeFramed {\advance\hsize-\width
   \@totalleftmargin\z@ \linewidth\hsize
   \@setminipage}}%
 {\par\unskip\endMakeFramed%
 \at@end@of@kframe}
\makeatother

\definecolor{shadecolor}{rgb}{.97, .97, .97}
\definecolor{messagecolor}{rgb}{0, 0, 0}
\definecolor{warningcolor}{rgb}{1, 0, 1}
\definecolor{errorcolor}{rgb}{1, 0, 0}
\newenvironment{knitrout}{}{} % an empty environment to be redefined in TeX

\usepackage{alltt}

\usepackage{orcidlink,thumbpdf,lmodern}
\usepackage{graphicx}
\usepackage{multirow}
\usepackage{multicol}
\usepackage{amsmath,amssymb,amsfonts, amsthm}
\usepackage[english]{babel}
\usepackage{booktabs}
\usepackage{pifont}
\usepackage{enumerate}
\usepackage{caption}
\usepackage{subcaption}
\usepackage{xcolor}
\usepackage{float}
\usepackage{threeparttable}
\usepackage{tabularx}

\newcommand{\fct}[1]{\code{#1()}}

\author{Joris Chau\\Open Analytics
\And Alexandre Segers\\Johnson \& Johnson\\Statistics and Decision Sciences
\AND Martin Otava\\Johnson \& Johnson\\Statistics and Decision Sciences
\And Stephen Karanja\\Johnson \& Johnson\\Statistics and Decision Sciences}
\Plainauthor{Joris Chau, Alexandre Segers, Martin Otava, Stephen Karanja}

\title{\pkg{cvxoptdes}: convex optimal experiment design in \proglang{R}}
\Plaintitle{cvxoptdes: convex optimal experiment design in R}
\Shorttitle{cvxoptdes: convex optimal experiment design in R}

\Abstract{
The \pkg{cvxoptdes} \proglang{R}-package provides a unified framework to perform convex optimal experiment design and online response optimization for linear, generalized linear and additive, nonlinear, and random intercept mixed models, using exclusively open-source solvers. It supports optimization and augmentation of weighted, penalized and convex constrained continuous designs across a discrete grid of candidate points and rounding to integer-valued designs using point-exchange algorithms or mixed-integer programming for orthogonally constrained designs, including the class of orthogonal minimally aliased response surface (OMARS) designs. Models and design constraints are defined and updated in an object-oriented context with the aim to facilitate the process of sequential (penalized) design augmentation analogous to online design procedures known from active learning and Bayesian optimization.
}

\Plainkeywords{cvxoptdes, DoE, Optimal design, Model-based optimization, Minimal aliasing, SOCP, R}

\Address{
  Joris Chau\\
  Open Analytics NV\\
  Jupiterstraat 20\\
  2600 Antwerp\\
  Belgium\\
  E-mail: \email{joris.chau@openanalytics.eu}
}
\IfFileExists{upquote.sty}{\usepackage{upquote}}{}
\begin{document}

%% -- Introduction -------------------------------------------------------------

%% - In principle "as usual".
%% - But should typically have some discussion of both _software_ and _methods_.
%% - Use \proglang{}, \pkg{}, and \code{} markup throughout the manuscript.
%% - If such markup is in (sub)section titles, a plain text version has to be
%%   added as well.
%% - All software mentioned should be properly \cite-d.
%% - All abbreviations should be introduced.
%% - Unless the expansions of abbreviations are proper names (like "Journal
%%   of Statistical Software" above) they should be in sentence case (like
%%   "generalized linear models" below).

\section[Introduction]{Introduction} \label{sec:1}

%% -- Manuscript ---------------------------------------------------------------

%% - In principle "as usual" again.
%% - When using equations (e.g., {equation}, {eqnarray}, {align}, etc.
%%   avoid empty lines before and after the equation (which would signal a new
%%   paragraph.
%% - When describing longer chunks of code that are _not_ meant for execution
%%   (e.g., a function synopsis or list of arguments), the environment {Code}
%%   is recommended. Alternatively, a plain {verbatim} can also be used.
%%   (For executed code see the next section.)

Experimental design plays a central role in industrial process development, where the goal is to characterize the relationship between process factors and outputs while making efficient use of available resources. Unlike model-free approaches such as factorial or space-filling designs, optimal designs select experimental conditions by optimizing a design criterion defined with respect to a specified candidate model. Common criteria include minimizing estimation uncertainty (D-/A-optimality), minimizing prediction variance (I-/G-optimality), and reducing aliasing among model terms, among others. Depending on the choice of model and optimality criterion, optimal designs can be used for a variety of goals, including factor screening, parameter estimation, prediction, and hypothesis testing.

Traditionally, optimal designs are generated in an offline setting: the entire design is constructed before any data are collected, and although it may occasionally be augmented afterward, such augmentations are usually limited in number. The objective of the design is to gain information across the whole experimental region so that reliable model predictions can be made at any setting of interest. On the other hand, in online design frameworks like Bayesian optimization \citep{S12}, the goal shifts from characterizing the entire response surface to locating the factor levels that optimize (e.g., maximize) the response. Starting from a small initial design, one or a few design points are proposed per iteration based on the current model predictions, balancing exploration of the design space against exploitation of promising regions to avoid local optima. Bayesian optimization, as well as active learning \citep{S09}, typically rely on nonparametric modeling, most notably Gaussian process regression, to estimate both the response and its associated uncertainty. Because the objective is targeted optimization instead of global characterization, these techniques can often identify near-optimal settings using substantially fewer experiments than classical offline optimal designs. However, in many industrial applications, response optimization is not the only requirement; parameter estimability, a well-understood variance structure, controlled prediction variance, and model interpretability are often equally important, yet these considerations are rarely the primary focus of online optimization methods. 

To reconcile these two aims, classical optimal design and sequential model-based optimization, within a single open-source framework, we have developed the \pkg{cvxoptdes} \proglang{R}-package. \pkg{cvxoptdes} allows for the optimization and augmentation of weighted, penalized, and convex constrained approximate designs across a discrete grid of candidate points for a variety of models. Penalization in the design criterion provides an explicit mechanism for balancing exploration and optimization: optimizing the design criterion alone drives exploration of the design space, while penalties informed by the current model's predictions shift emphasis toward design points expected to yield more favorable responses. Computationally, \pkg{cvxoptdes} solves for approximate optimal designs with continuous design weights via a second-order cone program (SOCP) using the Splitting Cone Solver \citep{SCS}. It then rounds the resulting approximate design to an exact integer-valued design using point-exchange algorithms, or, for orthogonally constrained designs, via mixed-integer programming with the HiGHS solver \citep{HIGHS}. 

\begin{figure}[h]
\center
\includegraphics[width=1\linewidth]{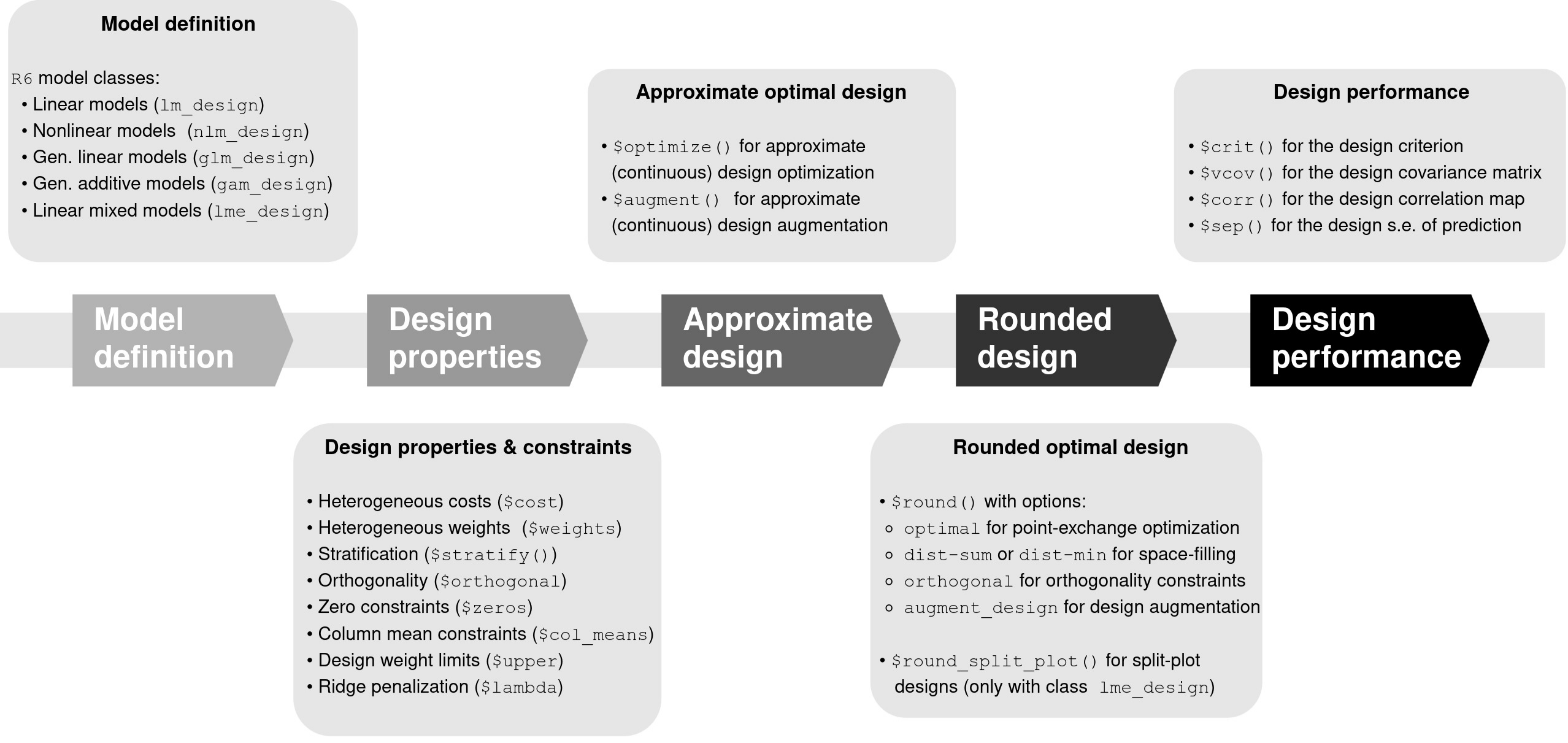}
\caption{High-level diagram of the \pkg{cvxoptdes} \proglang{R}-package. All class methods ending in parentheses (e.g., \fct{\$optimize}) are called as functions. All data fields (e.g., \code{\$cost}) are populated when initializing an \pkg{R6} model object.} \label{fig:diagram}
\end{figure}

\pkg{cvxoptdes} features an object-oriented interface based on \pkg{R6} \citep{R6}; a high-level workflow diagram of the exposed classes and class methods is provided in Figure \ref{fig:diagram}. Among its capabilities, the package supports:
\begin{itemize}
\item Approximate and exact D-, A-, I-, G-, and alias-optimal designs for linear models.
\item Locally optimal designs for nonlinear models and generalized linear and additive models.
\item Split-plot and stratified designs for random block effects models.
\item Orthogonality and other design constraints, enabling the computation of Orthogonal Minimally Aliased Response Surface \citep{NG20} designs. 
\item Heterogeneous grid weights or costs, enabling an explicit trade-off between design space exploration and response optimization.
\end{itemize}

Existing software packages for experimental design vary significantly in their supported features and underlying optimization algorithms. Prior to the development of \pkg{cvxoptdes}, multiple packages would have to be combined to cover the supported models and design types (Table \ref{tab:comparison}). As noted above, \pkg{cvxoptdes} uses the Splitting Cone Solver (SCS) to optimize approximate designs formulated as convex SOCPs. While this may be less efficient for exceptionally large design matrices compared to the gradient-ascent Fedorov algorithm in \pkg{AlgDesign} \citep{algdesign}, the SOCP framework offers a principal advantage: it allows the design problem to be extended directly with additional convex cone constraints, such as cost penalization, stratification, and orthogonality. Furthermore, recent advances in convex optimization \citep{CB25} offer significant solver accelerations. \pkg{OptimalDesign} \citep{optimaldesign} is also able to calculate approximate, as well as exact, optimal designs as SOCPs using the \pkg{Gurobi} solver, though this requires a commercial license for non-academic use. For exact optimal designs, \pkg{AlgDesign} generates solutions by rounding approximate designs \citep{PR92} or via a custom implementation of Fedorov's point-exchange algorithm, an approach similar to that of \pkg{cvxoptdes}. Similarly, \pkg{skpr} \citep{MK21} offers an efficient custom Fedorov point-exchange algorithm with optional parallelization. Commercial alternatives include \proglang{JMP}'s Custom Design interface, which relies on the coordinate-exchange algorithm \cite{MN95}, and \proglang{SAS}, which implements various exchange algorithms (e.g., \cite{MM70}; \cite{M74}; \cite{F72}).  

Beyond algorithmic differences, these packages also vary broadly in scope. \pkg{AlgDesign} supports approximate and exact optimal designs for linear models, optimal design augmentation, and blocked and split-plot designs with fixed block effects. While \pkg{skpr} does not generate approximate designs, it supports exact design optimization and augmentation across numerous optimality criteria, accommodates blocked and split-plot designs with random block effects, and provides a generic framework to evaluate the power of a design. Both \pkg{AlgDesign} and \pkg{skpr} use a stepwise approach for split-plot design generation, where the easy-to-change subplot levels are optimized conditional on the hard-to-change whole plot levels. \pkg{OptimalDesign} can handle locally optimal designs for specific nonlinear models (e.g., survival models), but does not generalize to formula-specified nonlinear models. \pkg{ICAOD} \citep{icaod} supports approximate locally, minimax, and Bayesian optimal designs for user-defined nonlinear models through a metaheuristic evolutionary algorithm, but does not natively calculate exact designs. Also, neither \pkg{OptimalDesign} nor \pkg{ICAOD} support blocked or split-plot designs. While both \pkg{cvxoptdes} and \proglang{JMP} offer many of the listed functionalities, only \pkg{cvxoptdes} provides them entirely within an open-source framework. Furthermore, \proglang{JMP} Pro features Bayesian optimization, aligning it with open-source tools like \pkg{mlr3mbo} \citep{mlr3mbo} or \pkg{BoTorch} \citep{botorch} for online response optimization. However, these tools primarily aim to discover optimal factor settings using data-driven non-parametric models, whereas \pkg{cvxoptdes} aims to integrate classical optimal design with response optimization. For many industrial applications, the parametric models supported by \pkg{cvxoptdes} are sufficient to model the response surface, offering greater interpretability and a reduced risk of overfitting in limited-data scenarios.

\begin{table}[t]
\centering
\resizebox{\textwidth}{!}{%
\begin{threeparttable}
\caption{Comparison of the scope and functionality of \proglang{R}-packages for optimal design. An open circle ($\circ$) indicates limited or partial support.} \label{tab:comparison}
\begin{tabular}{lcccc}
\toprule
 & \pkg{cvxoptdes} & \pkg{AlgDesign} & \pkg{skpr} & \pkg{OptimalDesign}\\
\midrule
Design criteria\tnote{*} & D, A, I, G, alias & D, A, I & D, A, I, G, T, E, Alias & D, A, I, C, c\\
Approximate design & $\checkmark$ & $\checkmark$ &  & $\checkmark$\\
Design augmentation & $\checkmark$ & $\checkmark$ & $\checkmark$ & $\checkmark$\\
Mixture designs & $\checkmark$ & $\checkmark$ & $\checkmark$ & $\checkmark$\\
Blocked and split-plot designs & $\checkmark$ & $\checkmark$\tnote{**} & $\checkmark$ & \\
Nonlinear models & $\checkmark$ &  &  & $\circ$\\
Generalized linear models & $\checkmark$ &  &  & $\circ$\\
Design penalization & $\checkmark$ &  &  & $\checkmark$\tnote{$\dagger$} \\
Orthogonality and stratification & $\checkmark$ &  &  & $\circ$\tnote{$\dagger$} \\
OMARS designs & $\checkmark$ &  &  & \\
Design power analysis &  &  & $\checkmark$ & \\
Open-source & $\checkmark$ & $\checkmark$ & $\checkmark$ & $\circ$\tnote{$\ddagger$} \\
\bottomrule
\end{tabular}
\begin{tablenotes}
\item \pkg{ICAOD} is not included in the comparison as its scope differs significantly from the other \proglang{R}-packages.
\item[*] For \pkg{AlgDesign} and \pkg{OptimalDesign}, the available design criteria depend on the function that is used. \pkg{skpr}'s Alias-optimality uses the trace criterion of \cite{JN11}, which is different from the minimal aliasing criterion in \pkg{cvxoptdes}.
\item[**] \pkg{AlgDesign} considers a fixed effects blocks model only, whereas \pkg{skpr} and \pkg{cvxoptdes} treat the blocks as random effects.     
\item[$\dagger$] \pkg{OptimalDesign} allows to constrain permissible designs by including linear equality and inequality constraints on the design weights.
\item[$\ddagger$] \pkg{OptimalDesign} requires \pkg{Gurobi} to solve integer quadratic and second-order cone programs. 
\end{tablenotes}
\end{threeparttable}%
}
\end{table}

The remainder of this paper is organized as follows. Section \ref{sec:2} introduces the class methods for approximate and exact optimal design of linear models, extending to penalized optimal designs, the inclusion of design constraints, and the generation of OMARS designs. Section \ref{sec:3} reviews additional model types, covering locally optimal designs for nonlinear and generalized linear models, as well as blocked and split-plot designs. Section \ref{sec:4} discusses optimal design augmentation and demonstrates how penalization can be used to achieve online design strategies similar to those in Bayesian optimization and active learning. Finally, Section \ref{sec:5} provides concluding remarks and potential future extensions.

\section{Linear model optimal design} \label{sec:2}

The \pkg{cvxoptdes} package adopts an objected-oriented framework to optimize or augment designs in order to reuse the defined model and grid of candidate design points across evaluations. The design object for a linear model is initialized using the \code{lm_design} \pkg{R6}-class:
\begin{knitrout}
\definecolor{shadecolor}{rgb}{1, 1, 1}\color{fgcolor}\begin{kframe}
\begin{alltt}
\hldef{> }\hlkwd{library}\hldef{(cvxoptdes)}
\hldef{> }
\hldef{> }\hldef{candidates} \hlkwb{<-} \hlkwd{expand.grid}\hldef{(}
\hldef{+ }  \hlkwc{x1} \hldef{=} \hlkwd{seq}\hldef{(}\hlkwc{from} \hldef{=} \hlopt{-}\hlnum{1}\hldef{,} \hlkwc{to} \hldef{=} \hlnum{1}\hldef{,} \hlkwc{length} \hldef{=} \hlnum{5}\hldef{),}
\hldef{+ }  \hlkwc{x2} \hldef{=} \hlkwd{seq}\hldef{(}\hlkwc{from} \hldef{=} \hlopt{-}\hlnum{1}\hldef{,} \hlkwc{to} \hldef{=} \hlnum{1}\hldef{,} \hlkwc{length} \hldef{=} \hlnum{5}\hldef{)}
\hldef{+ }\hldef{) |>}
\hldef{+ }  \hlkwd{subset}\hldef{(}\hlnum{2} \hlopt{*} \hldef{x1} \hlopt{+} \hldef{x2} \hlopt{> -}\hlnum{2.5} \hlopt{&} \hlnum{2} \hlopt{*} \hldef{x1} \hlopt{+} \hldef{x2} \hlopt{<} \hlnum{1.5}\hldef{)}
\hldef{> }
\hldef{> }\hldef{design} \hlkwb{<-} \hldef{lm_design}\hlopt{$}\hlkwd{new}\hldef{(}
\hldef{+ }  \hlkwc{formula} \hldef{=} \hlopt{~} \hldef{x1} \hlopt{*} \hldef{x2} \hlopt{+} \hlkwd{I}\hldef{(x1}\hlopt{^}\hlnum{2}\hldef{)} \hlopt{+} \hlkwd{I}\hldef{(x2}\hlopt{^}\hlnum{2}\hldef{),}
\hldef{+ }  \hlkwc{data} \hldef{= candidates}
\hldef{+ }\hldef{)}
\end{alltt}
\end{kframe}
\end{knitrout}
This example is taken from \cite[Example 12.2]{ADT07} and represents a simplified design problem in testing car engines assuming a second-order model in terms of the spark advance $x_1$ and the torque $x_2$, where certain combinations of the two factors are excluded to ensure the engine runs or to avoid damaging the engine. 

\subsection{Approximate optimal design} \label{sec:2.1}

After initialization of the design object, an approximate optimal design can be generated. By an approximate design, we refer to a design that is \emph{approximate} in the sense that a continuous weight is assigned to each candidate design point instead of an exact number of replicates. The latter is referred to as an exact or integer design throughout this work. To calculate an approximate optimal design for a linear model with $p$ coefficients across a grid of $n$ candidate design points, the \fct{\$optimize} method is used. This method solves the following penalized design optimization problem:
\begin{eqnarray}\label{eq:1}
\boldsymbol{w}^* = \arg \max_{\boldsymbol{w} \in \Delta^n} \Phi(M(\boldsymbol{w}\,|\,\boldsymbol{X},\boldsymbol{v}) + \lambda I) - \alpha \cdot \boldsymbol{c}^T \boldsymbol{w}
\end{eqnarray}
In this equation, $\boldsymbol{w} = (w_1,\ldots,w_n)^T$ is a vector of design weights across the candidate grid in the unit simplex $\Delta^n = \left\{ \boldsymbol{w} \in [0, 1]^n\,|\, \sum_{i=1}^n w_i = 1 \right\}$. $\boldsymbol{c} = (c_1,\ldots, c_n)^T$ are fixed costs assigned to each design weight scaled by a cost tuning parameter $\alpha$. $M(\boldsymbol{w}\,|\,\boldsymbol{X}, \boldsymbol{v})$ is the information matrix of the linear model depending on the weights $\boldsymbol{w}$, the $(n \times p)$-design matrix $\boldsymbol{X}$, and possibly a vector of fixed user weights $\boldsymbol{v} = (v_1,\ldots,v_n)^T$ in the case of weighted linear regression. Finally, $\Phi(M)$ is the design optimality criterion as a scalar function of the information matrix $M$.
Given the linear model \code{formula}, the $(p \times p)$-information matrix, conditional on the error variance $\sigma_e^2$ is given by:
\begin{eqnarray} \label{eq:2}
M(\boldsymbol{w}\,|\,\boldsymbol{X},\boldsymbol{v}) \ = \ \frac{1}{\sigma_e^2} \sum_{i = 1}^n w_i v_i \boldsymbol{x}_i \boldsymbol{x}_i^T
\end{eqnarray}
where $\boldsymbol{x}_i$ denotes the $i$-th row in the design matrix $\boldsymbol{X}$. If $\lambda > 0$ in the design optimization problem, a ridge term is added to the diagonal of $M$ resulting in the information matrix of a ridge linear regression model. This is particularly useful when $n < p$ to ensure a non-singular information matrix for the evaluation of the optimality criterion $\Phi(M)$.

The approximate optimal design problem is solved by the \fct{\$optimize} method in the form of a convex second-order cone program (SOCP) using the open-source Splitting Cone Solver \citep{SCS}. For D-optimal designs, the SOCP is formulated using spectral cones introduced in \cite{CB25}. For I-, A-, and G-optimal designs, the SOCP formulations of \cite{SH15} are used. The following optimality criteria are currently supported by \pkg{cvxoptdes}:

\begin{description}
\item[\code{D}-optimality]: Maximize the volume of the information matrix, or equivalently minimize the volume of the confidence ellipsoid of the parameter estimates. For D-optimality, the logarithm of the D-criterion is used for optimization: 
\begin{eqnarray}
\Phi(M) \ = \ \log\left(\det(M)^{1/p}\right)
\end{eqnarray} 
\item[\code{A}-optimality]: Maximize the inverse trace of the inverse information, or equivalently minimize the average variance of the parameter estimates:
\begin{eqnarray}
\Phi(M) \ = \ \text{Tr}(M^{-1})^{-1}
\end{eqnarray}
\item[\code{I}-optimality]: Maximize the inverse of the average prediction variance across the candidate design grid, (equivalently minimize the average prediction variance):
\begin{eqnarray}
\Phi(M) \ = \ \left(\frac{1}{n} \sum_{i = 1}^n \boldsymbol{x}_i' M^{-1} \boldsymbol{x}_i\right)^{-1}
\end{eqnarray}
\item[\code{G}-optimality]: Maximize the inverse of the maximum prediction variance across the candidate design grid, (equivalently minimize the maximum prediction variance):
\begin{eqnarray}
\Phi(M) = \left(\max_{\boldsymbol{x}_i} \boldsymbol{x}_i' M^{-1} \boldsymbol{x}_i\right)^{-1}
\end{eqnarray}
\item[\code{alias}-optimality]: Maximize one minus the average absolute cosine similarity
(also the uncentered correlation or the normalized dot product) between the columns of
the design matrix:
\begin{eqnarray}
\Phi(M) \ = \ 1 - \frac{2}{p(p - 1)} \sum_{1 \leq q < r \leq p} \left| \frac{M_{rq}}{\sqrt{M_{qq}}\sqrt{M_{rr}}} \right|
\end{eqnarray}
If the weighted design columns are all centered in the sense that $\sum_i w_i x_{iq} = 0$ for $q = 1,\ldots, p$, this criterion is equivalent to minimizing the average absolute correlation between the design columns.
\end{description}

For the car engine example, the standard approximate D-optimal design is evaluated by:
\begin{knitrout}
\definecolor{shadecolor}{rgb}{1, 1, 1}\color{fgcolor}\begin{kframe}
\begin{alltt}
\hldef{> }\hldef{design}\hlopt{$}\hlkwd{optimize}\hldef{(}\hlkwc{criterion} \hldef{=} \hlsng{"D"}\hldef{) |>} \hlkwd{round}\hldef{(}\hlnum{3}\hldef{)}
\end{alltt}
\begin{verbatim}
 [1] 0.122 0.096 0.000 0.162 0.000 0.000 0.000 0.140 0.000 0.129 0.039 0.000
[13] 0.000 0.000 0.156 0.000 0.156
attr(,"criterion")
attr(,"criterion")$crit.value
[1] 0.2863781

attr(,"criterion")$crit
[1] "D"
\end{verbatim}
\end{kframe}
\end{knitrout}
The returned design weights are in the same order as the \code{data} grid of candidate points and the vector includes the optimized criterion value as an attribute. 

\subsection{Exact optimal design}

To obtain an exact design with an integer number of replicates per design point, the \fct{\$round} method is used. The default rounding method is \code{optimal} rounding, which executes a number (\code{n_repeats}) of repeated Fedorov point-exchange procedures \citep{F72}, where each greedy exchange loop runs until no improving point exchange exists or when the maximum number of exchange iterations (\code{max_iter}) is reached. The point-exchange procedures start from randomly sampled versions of the distribution of approximate design weights, or from \emph{efficiently rounded} designs \citep{PR92} if the size of the exact design is sufficiently large. 

\begin{figure}[t]
\center
\includegraphics[scale=1]{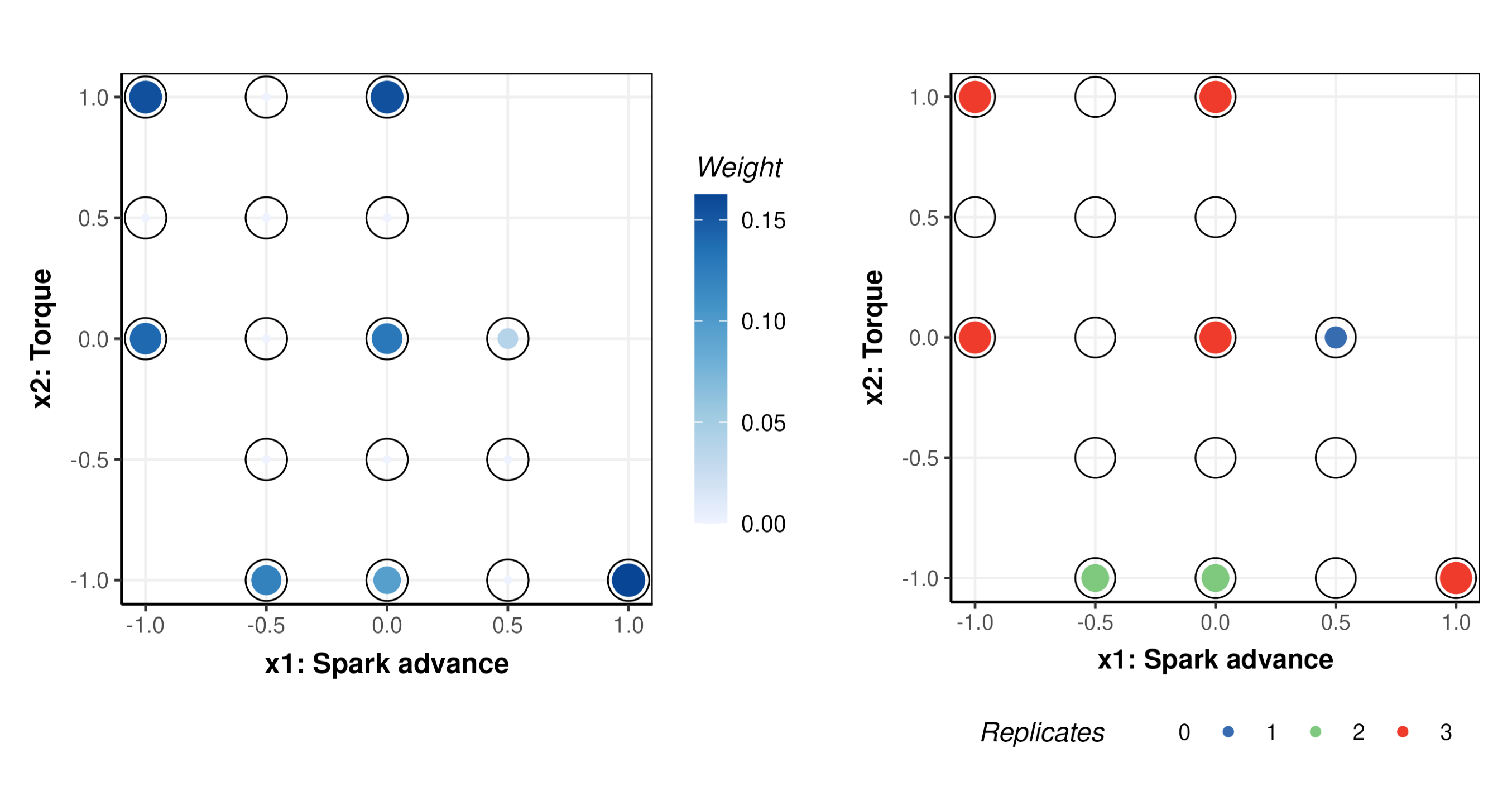}
\caption{Car engine example approximate D-optimal design (left) and exact D-optimal design of size $m = 20$ (right). Outer circles represent candidate design points, inner circles represent selected design points colored/sized by weight (left) or by replicate count (right).} \label{fig:1}
\end{figure}

Because the starting design is informed by the continuous optimal design, it is expected that the exchange algorithm generally requires less iterations to reach the globally optimal exact design than initializing the exchange procedure from a random integer design, which is the default approach in most software packages, (e.g. \proglang{SAS OPTEX}, \proglang{JMP Custom Design} or \pkg{skpr}). This is especially true if the design size is large, since the exact and approximate optimal design converge as demonstrated in Figure \ref{fig:1}, whereas it becomes increasingly difficult to identify the optimal design starting from a random initial design due to the curse of dimensionality. The exchange procedure initially only swaps candidate points with nonzero support in the approximate optimal design. If the design can no longer be improved by exchanging points from the restricted candidate grid, the exchanges are extended to the complete grid of candidate design points. This two-stage approach is most useful when the number of candidate design points is large relative to the support size of the approximate design, as it provides a good heuristic in selecting candidate points that are relevant to build the exact design. The computationally expensive exchange criterion evaluations in the point-exchange algorithm are already optimized for efficiency using low-level \code{BLAS} (and \code{LAPACK}) routines, but additional performance improvements are possible by executing multiple exchange loops in parallel with the \pkg{mirai} \proglang{R}-package \citep{mirai}.

An exact D-optimal design with $m = 20$ design points for the car engine example is generated as follows:
\begin{knitrout}
\definecolor{shadecolor}{rgb}{1, 1, 1}\color{fgcolor}\begin{kframe}
\begin{alltt}
\hldef{> }\hldef{design}\hlopt{$}\hlkwd{round}\hldef{(}\hlkwc{m} \hldef{=} \hlnum{20}\hldef{,} \hlkwc{method} \hldef{=} \hlsng{"optimal"}\hldef{,} \hlkwc{n_repeats} \hldef{=} \hlnum{10}\hldef{,} \hlkwc{max_iter} \hldef{=} \hlnum{100}\hldef{)}
\end{alltt}
\begin{verbatim}
 [1] 2 2 0 3 0 0 0 3 0 3 1 0 0 0 3 0 3
attr(,"criterion")
attr(,"criterion")$crit
[1] "D"

attr(,"criterion")$crit.value
[1] 0.2851871

attr(,"criterion")$crit.rel.eff
[1] 0.9958412
\end{verbatim}
\end{kframe}
\end{knitrout}
The returned vector holds the number of replicates per design point in the same order as the \code{data} candidate grid. In addition to the optimized criterion value of the exact design, the relative efficiency with respect to the approximate optimal design is included, where the criterion value of the approximate optimal design provides an upper limit for the criterion value of any finite exact design. Given the integer replicate vector, the selected design points can be retrieved by subsetting the \code{data} grid or by using the dedicated \fct{\$subset} method. The returned exact design matches the D-optimal design obtained in \cite[Example 12.2]{ADT07} and is displayed in Figure \ref{fig:1}.
\subsection{Penalized optimal design} \label{sec:2.3}

By penalized or weighted optimal design, we refer to optimization of the design criterion subject to a user-defined cost penalty or weight assigned to each candidate design point. In a chemical or pharmaceutical setting, performing experiments at standard room temperature, humidity and atmospheric pressure may be less costly and complex than operating under highly stressed conditions. Also, chemical compounds, such as solvents or reactants in a chemical reaction, may vary in terms of safety, availability and price, or other chemical properties that are preferable for the scientist performing the experiment. Another important use-case of cost penalization is in the context of global optimization of a response, e.g. chemical yield or purity, through sequential design augmentation. In this case the cost penalties are taken (inversely) proportional to the predictions of the response surface across the candidate design grid, thereby promoting candidate points with a high predicted response value when augmenting the design. This is similar to the exploration versus exploitation paradigm in Bayesian Optimization and the topic of penalized design augmentation is treated in detail in Section \ref{sec:4.1}.

Heterogeneous weights or costs can be included in the design optimization problem in two ways: through the weight vector $\boldsymbol{v} = (v_1,\ldots,v_n)^T$, or through the cost vector $\boldsymbol{c} = (c_1,\ldots,c_n)^T$. The weights $\boldsymbol{v}$ are incorporated as multiplicative terms at the level of the information matrix, whereas the costs $\boldsymbol{c}$ are penalizing the optimality criterion value using an additive penalty term. Both options have intuitive interpretations; a heterogeneous weight vector $\boldsymbol{v}$ produces the optimal design of a weighted linear regression model. For a non-zero cost vector $\boldsymbol{c}$, the problem in eq.(\ref{eq:1}) is the dual form (with Lagrange multiplier $\alpha$) of the cost-constrained design optimization problem:
\begin{eqnarray}
\boldsymbol{w}^* = \arg \max_{\boldsymbol{w} \in \Delta^n} \Phi(M(\boldsymbol{w}\,|\,\boldsymbol{X},\boldsymbol{v}) + \lambda I),\quad \quad \text{subject to}\quad \boldsymbol{c}^T \boldsymbol{w} \leq B
\end{eqnarray}
with design budget $B$. An advantage of using the multiplicative weights $\boldsymbol{v}$ is that the weighted optimal design is invariant under scaling of $\boldsymbol{v}$. This is not the case when scaling the penalty parameter $\alpha$ or the costs $\boldsymbol{c}$, which change the impact of the additive penalty term on the overall optimization objective. 

\subsubsection{Imidazoles reaction example}

\begin{figure}[t]
\center
\includegraphics[width=1\linewidth]{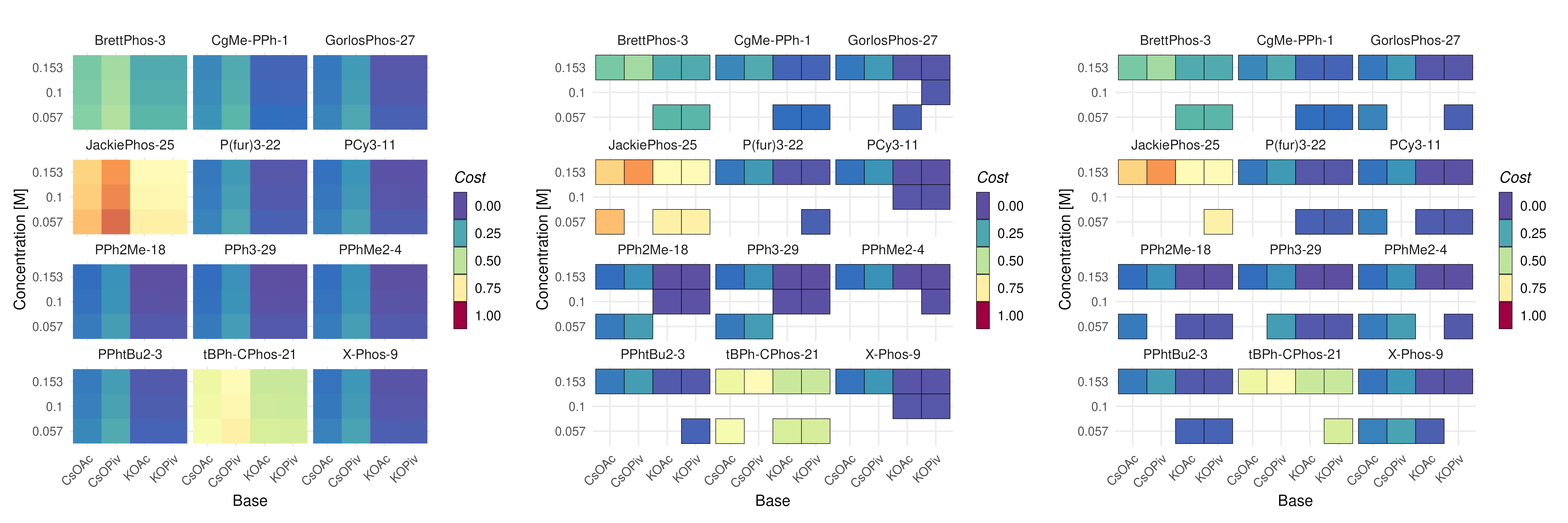}
\caption{Imidazoles reaction candidate grid colored by cost with ligands as facets (left); $\boldsymbol{v}$-weighted D-optimal design of size $m = 75$ (middle); and $\boldsymbol{c}$-penalized D-optimal design of size $m = 75$ with $\alpha = 1$ (right).} \label{fig:2}
\end{figure}

In the context of chemical reaction optimization, the \code{imidazoles} dataset (\cite{S21}, \cite{T22}) reports experimental yields and reaction costs in a direct arylation reaction of imidazoles measured across all 1728 combinations of 3 temperatures, 3 initial substrate concentrations, 4 base molecules, 4 solvent molecules and 12 ligand molecules. For demonstration, only 144 experiments at the single solvent DMAc and temperature 105°C are kept. To generate an optimal design, we impose a linear model with all main-effects and first-order interaction terms for the discrete concentration and categorical base and ligand variables (63 model parameters). Using a weight vector $\boldsymbol{v}$ equal to the inverse squared reaction costs, the \pkg{R6}-design object is initialized as follows:
\begin{knitrout}
\definecolor{shadecolor}{rgb}{1, 1, 1}\color{fgcolor}\begin{kframe}
\begin{alltt}
\hldef{> }\hldef{design} \hlkwb{<-} \hldef{lm_design}\hlopt{$}\hlkwd{new}\hldef{(}
\hldef{+ }  \hlkwc{formula} \hldef{=} \hlopt{~} \hldef{(base} \hlopt{+} \hldef{ligand} \hlopt{+} \hldef{concentration)}\hlopt{^}\hlnum{2}\hldef{,}
\hldef{+ }  \hlkwc{weights} \hldef{=} \hlopt{~} \hlnum{1} \hlopt{/} \hldef{cost}\hlopt{^}\hlnum{2}\hldef{,}
\hldef{+ }  \hlkwc{data} \hldef{=} \hlkwd{subset}\hldef{(imidazoles, solvent} \hlopt{==} \hlsng{"DMAc"} \hlopt{&} \hldef{temperature} \hlopt{==} \hlnum{105}\hldef{)}
\hldef{+ }\hldef{)}
\end{alltt}
\end{kframe}
\end{knitrout}
A weighted D-optimal design of size $m=75$ is displayed in Figure \ref{fig:2} (middle), optimized in the same way as for the car engine example using \code{optimal} rounding for the exact design. For penalization using a cost vector $\boldsymbol{c}$, the existing object can be updated with the \fct{\$update} method. 
\begin{knitrout}
\definecolor{shadecolor}{rgb}{1, 1, 1}\color{fgcolor}\begin{kframe}
\begin{alltt}
\hldef{> }\hldef{design}\hlopt{$}\hlkwd{update}\hldef{(}
\hldef{+ }  \hlkwc{weights} \hldef{=} \hlkwa{NULL}\hldef{,}
\hldef{+ }  \hlkwc{cost} \hldef{=} \hlopt{~}\hldef{cost,}
\hldef{+ }  \hlkwc{alpha} \hldef{=} \hlnum{1}
\hldef{+ }\hldef{)}
\end{alltt}
\end{kframe}
\end{knitrout}
In Figure \ref{fig:2} on the right, a penalized D-optimal design of the same size ($m=75$) as the weighted D-optimal design is shown. Both designs behave similarly in biasing the proposed design points towards the low-cost regions of the candidate grid, but the two approaches are fundamentally different in how candidate points are penalized in the optimization problem.

\subsection{Design constraints}

\subsubsection{Stratified optimal design}

The formulation of the approximate design optimization problem as a convex second-order cone program allows for straightforward extension by extra convex cone constraints. One of such constraints is that of a stratified optimal design, which partitions the candidate design grid into individual strata $\mathcal{X}_1,\ldots,\mathcal{X}_S$ and solves the optimal design problem requiring that each stratum is represented with a given proportion $p_1,\ldots,p_S$:
\begin{eqnarray}
\boldsymbol{w}^* & = &  \arg \max_{\boldsymbol{w} \in \Delta^n} \Phi(M(\boldsymbol{w}\,|\,\boldsymbol{X},\boldsymbol{v}) + \lambda I) - \alpha \cdot \boldsymbol{c}^T \boldsymbol{w}, \\
&& \text{subject to } \sum_{\{i:\,\boldsymbol{x}_i \in \mathcal{X}_s\}} w_i = p_s, \quad \text{for } s = 1,\ldots,S
\end{eqnarray}
To include stratification constraints, the \fct{\$stratify} method is used. In the \code{imidazoles} example, we can require that each ligand is represented equally in the approximate and exact optimal design with:
\begin{knitrout}
\definecolor{shadecolor}{rgb}{1, 1, 1}\color{fgcolor}\begin{kframe}
\begin{alltt}
\hldef{> }\hldef{design}\hlopt{$}\hlkwd{stratify}\hldef{(}
\hldef{+ }  \hlkwc{strata} \hldef{=} \hlopt{~}\hldef{ligand,}
\hldef{+ }  \hlkwc{proportions} \hldef{=} \hlkwd{rep}\hldef{(}\hlnum{1} \hlopt{/} \hlnum{12}\hldef{,} \hlnum{12}\hldef{)}
\hldef{+ }\hldef{)}
\end{alltt}
\end{kframe}
\end{knitrout}
This type of stratification only defines feasibility constraints on the allocation of the design weights, for instance because equal representation of certain subgroups or subpopulations is desired in the generated design. The linear model error structure and associated information matrix are not changed and should not be confused with the construction of \emph{blocked} designs, where the block effects are included in the model as nuisance random or fixed effects \citep[Chapter 4]{G02}.

\subsubsection{Design orthogonality constraints}

The \code{alias}-optimality criterion minimizes the absolute cosine
similarity between columns of the design matrix $\boldsymbol{X}$. Instead of minimizing the
cosine similarity, it is also possible to add explicit constraints on the absolute cosine
similarities, which are representable as (convex) rotated second-order cone
inequalities. For design columns $q$ and $r$, and a given constraint
value $\rho_{qr} \in [0, 1]$, we can include the following orthogonality constraint in the approximate design optimization problem:
\begin{eqnarray} \label{eq:10}
\left| \frac{\langle \boldsymbol{w} \boldsymbol{x}_{.q}, \boldsymbol{x}_{.r} \rangle}{\Vert \sqrt{\boldsymbol{w}} \boldsymbol{x}_{.q} \Vert \Vert \sqrt{\boldsymbol{w}} \boldsymbol{x}_{.r} \Vert} \right| \ \leq \ \rho_{qr}
\end{eqnarray}
Returning to the car engine example, we can print the correlation matrix between the design columns subject to the approximate D-optimal design calculated at the start of this section with the \fct{\$corr} method. It can be verified that none of the design columns are uncorrelated under the approximate optimal design:
\begin{knitrout}
\definecolor{shadecolor}{rgb}{1, 1, 1}\color{fgcolor}\begin{kframe}
\begin{alltt}
\hldef{> }\hldef{design}\hlopt{$}\hlkwd{corr}\hldef{(}
\hldef{+ }  \hlkwc{design_weights} \hldef{=} \hlkwd{c}\hldef{(}
\hldef{+ }    \hlnum{0.122}\hldef{,} \hlnum{0.0956}\hldef{,} \hlnum{0}\hldef{,} \hlnum{0.162}\hldef{,} \hlnum{0}\hldef{,} \hlnum{0}\hldef{,} \hlnum{0}\hldef{,} \hlnum{0.140}\hldef{,} \hlnum{0}\hldef{,}
\hldef{+ }    \hlnum{0.129}\hldef{,} \hlnum{0.039}\hldef{,} \hlnum{0}\hldef{,} \hlnum{0}\hldef{,} \hlnum{0}\hldef{,} \hlnum{0.156}\hldef{,} \hlnum{0}\hldef{,} \hlnum{0.156}
\hldef{+ }  \hldef{)}
\hldef{+ }\hldef{)}
\end{alltt}
\end{kframe}
\end{knitrout}
To reduce the correlation between the design columns, the \code{orthogonal} argument in \fct{\$new} or \fct{\$update} is used, which accepts a three-part formula, e.g. \code{x1 + x2 ~ x1 + x2 ~ 0}. The first part (\code{x1 + x2}) defines all design columns $\boldsymbol{x}_{.q}$ in eq.(\ref{eq:10}); the second part (\code{x1 + x2}) defines all design columns $\boldsymbol{x}_{.r}$ in eq.(\ref{eq:10}); the third part is optional and specifies the constraint value $\rho_{qr}$. If only a two-part formula is provided, $\rho_{qr}$ is set to zero. 
\begin{knitrout}
\definecolor{shadecolor}{rgb}{1, 1, 1}\color{fgcolor}\begin{kframe}
\begin{alltt}
\hldef{> }\hldef{design}\hlopt{$}\hlkwd{update}\hldef{(}
\hldef{+ }  \hlkwc{orthogonal} \hldef{= x1} \hlopt{+} \hldef{x2} \hlopt{~} \hldef{x1} \hlopt{+} \hldef{x2}
\hldef{+ }\hldef{)}
\hldef{> }\hldef{design}\hlopt{$}\hlkwd{optimize}\hldef{(}\hlkwc{criterion} \hldef{=} \hlsng{"D"}\hldef{) |>} \hlkwd{round}\hldef{(}\hlnum{3}\hldef{)}
\end{alltt}
\begin{verbatim}
 [1] 0.174 0.055 0.000 0.059 0.000 0.000 0.000 0.064 0.000 0.118 0.241 0.000
[13] 0.000 0.000 0.028 0.000 0.260
attr(,"criterion")
attr(,"criterion")$crit.value
[1] 0.2208041

attr(,"criterion")$crit
[1] "D"
\end{verbatim}
\end{kframe}
\end{knitrout}
The additional orthogonality constraint results in a sub-optimal design in terms of D-optimality, but the main effects columns have become uncorrelated in the constrained approximate optimal design.

For exact designs, the orthogonality constraints specified in the \code{orthogonal} argument are not enforced by the \code{optimal} rounding procedure. To round an orthogonally constrained approximate design, the dedicated \code{orthogonal} rounding procedure should be used instead, which attempts to find a feasible integer design that satisfies the imposed orthogonality constraints for a given integer number of design points. This is done by solving a mixed-integer linear problem (MILP) using the open-source HiGHS solver \citep{HIGHS}. Since the cosine similarity constraint in eq.(\ref{eq:10}) cannot be simplified to an exact linear constraint as required for an MILP, the design weights in the denominator needed to standardize the absolute inner product are substituted by the fixed weights obtained from the approximate design. This means that for nonzero $\rho_{qr}$, the orthogonality constraints in the MILP are not exactly equal to the constraints in eq.(\ref{eq:10}). When  $\rho_{qr} = 0$, the scaling of the inner product can be dropped, so this problem disappears for exact orthogonality constraints. 

\subsubsection{Reactor experiment example}

\begin{figure}[t]
\center
\includegraphics[width=1\linewidth]{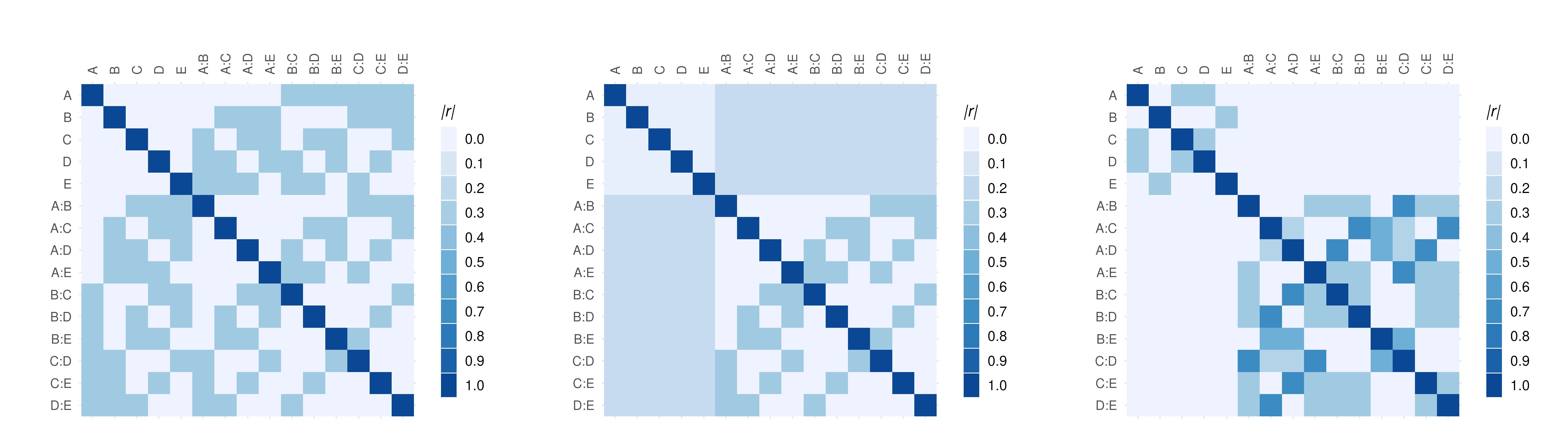}
\caption{Reactor experiment correlation maps of 12-run orthogonally constrained designs. From left to right: Plackett-Burman; Bayesian D-optimal; D-efficient minimal aliasing.} \label{fig:3}
\end{figure}

We illustrate the generation of several exact designs with orthogonality constraints by replicating the reactor experiment example from \cite{BHH05} and \cite{JN11}. The candidate design grid consists of all $2^5$ combinations of five two-level factors: Rate (A), Catalyst (B), Stir Rate (C), Temperature (D), and Concentration (E). The objective is to construct a small screening design based on a linear model including the main effects and two-factor interactions:
\begin{knitrout}
\definecolor{shadecolor}{rgb}{1, 1, 1}\color{fgcolor}\begin{kframe}
\begin{alltt}
\hldef{> }\hldef{candidates} \hlkwb{<-} \hlkwd{expand.grid}\hldef{(}
\hldef{+ }  \hlkwc{A} \hldef{=} \hlkwd{c}\hldef{(}\hlopt{-}\hlnum{1}\hldef{,} \hlnum{1}\hldef{),}
\hldef{+ }  \hlkwc{B} \hldef{=} \hlkwd{c}\hldef{(}\hlopt{-}\hlnum{1}\hldef{,} \hlnum{1}\hldef{),}
\hldef{+ }  \hlkwc{C} \hldef{=} \hlkwd{c}\hldef{(}\hlopt{-}\hlnum{1}\hldef{,} \hlnum{1}\hldef{),}
\hldef{+ }  \hlkwc{D} \hldef{=} \hlkwd{c}\hldef{(}\hlopt{-}\hlnum{1}\hldef{,} \hlnum{1}\hldef{),}
\hldef{+ }  \hlkwc{E} \hldef{=} \hlkwd{c}\hldef{(}\hlopt{-}\hlnum{1}\hldef{,} \hlnum{1}\hldef{)}
\hldef{+ }\hldef{)}
\hldef{> }\hldef{design} \hlkwb{<-} \hldef{lm_design}\hlopt{$}\hlkwd{new}\hldef{(}
\hldef{+ }  \hlkwc{formula} \hldef{=} \hlopt{~} \hldef{(A} \hlopt{+} \hldef{B} \hlopt{+} \hldef{C} \hlopt{+} \hldef{D} \hlopt{+} \hldef{E)}\hlopt{^}\hlnum{2}\hldef{,}
\hldef{+ }  \hlkwc{data} \hldef{= candidates}
\hldef{+ }\hldef{)}
\end{alltt}
\end{kframe}
\end{knitrout}
The first design is a 12-run design constraining all main effects to be uncorrelated. The correlation map of the exact design shown on the left in Figure \ref{fig:3} turns out to be identical to that of the Plackett-Burman design in \cite{JN11}.
\begin{knitrout}
\definecolor{shadecolor}{rgb}{1, 1, 1}\color{fgcolor}\begin{kframe}
\begin{alltt}
\hldef{> }\hldef{design}\hlopt{$}\hlkwd{update}\hldef{(}
\hldef{+ }  \hlkwc{orthogonal} \hldef{= A} \hlopt{+} \hldef{B} \hlopt{+} \hldef{C} \hlopt{+} \hldef{D} \hlopt{+} \hldef{E} \hlopt{~} \hldef{A} \hlopt{+} \hldef{B} \hlopt{+} \hldef{C} \hlopt{+} \hldef{D} \hlopt{+} \hldef{E}
\hldef{+ }\hldef{)}
\hldef{> }\hldef{design}\hlopt{$}\hlkwd{optimize}\hldef{(}\hlkwc{criterion} \hldef{=} \hlsng{"D"}\hldef{)}
\hldef{> }\hldef{design}\hlopt{$}\hlkwd{round}\hldef{(}\hlkwc{m} \hldef{=} \hlnum{12}\hldef{,} \hlkwc{method} \hldef{=} \hlsng{"orthogonal"}\hldef{)}
\end{alltt}
\end{kframe}
\end{knitrout}
The second design does not impose exact orthogonality, but instead limits the absolute cosine similarity between the main effects columns and all other design columns to $\rho_{qr} = 0.3$. The correlation map is shown in the middle in Figure \ref{fig:3} and coincides with that of the Bayesian D-optimal design generated in \cite{JN11}. 
\begin{knitrout}
\definecolor{shadecolor}{rgb}{1, 1, 1}\color{fgcolor}\begin{kframe}
\begin{alltt}
\hldef{> }\hldef{design}\hlopt{$}\hlkwd{update}\hldef{(}
\hldef{+ }  \hlkwc{orthogonal} \hldef{= A} \hlopt{+} \hldef{B} \hlopt{+} \hldef{C} \hlopt{+} \hldef{D} \hlopt{+} \hldef{E} \hlopt{~} \hldef{.} \hlopt{~} \hlnum{0.3}
\hldef{+ }\hldef{)}
\hldef{> }\hldef{design}\hlopt{$}\hlkwd{round}\hldef{(}\hlkwc{m} \hldef{=} \hlnum{12}\hldef{,} \hlkwc{method} \hldef{=} \hlsng{"orthogonal"}\hldef{)}
\end{alltt}
\end{kframe}
\end{knitrout}
The third design combines multiple constraint formulas in the \code{orthogonal} argument. The first formula limits the absolute cosine similarities between all design columns to $\rho_{qr} = 0.75$. In addition, the second formula constrains the main effects columns to be exactly orthogonal to all interaction columns. The correlation map is shown on the right in Figure \ref{fig:3}. The generated design is not identical to the D-efficient minimal aliasing design in \cite{JN11}, but is equivalent in terms of the total (summed) amount of correlation between the main effects columns and between the interaction columns.
\begin{knitrout}
\definecolor{shadecolor}{rgb}{1, 1, 1}\color{fgcolor}\begin{kframe}
\begin{alltt}
\hldef{> }\hldef{design}\hlopt{$}\hlkwd{update}\hldef{(}
\hldef{+ }  \hlkwc{orthogonal} \hldef{=} \hlkwd{list}\hldef{(}
\hldef{+ }    \hldef{(A} \hlopt{+} \hldef{B} \hlopt{+} \hldef{C} \hlopt{+} \hldef{D} \hlopt{+} \hldef{E)}\hlopt{^}\hlnum{2} \hlopt{~} \hldef{(A} \hlopt{+} \hldef{B} \hlopt{+} \hldef{C} \hlopt{+} \hldef{D} \hlopt{+} \hldef{E)}\hlopt{^}\hlnum{2} \hlopt{~} \hlnum{0.75}\hldef{,}
\hldef{+ }    \hldef{A} \hlopt{+} \hldef{B} \hlopt{+} \hldef{C} \hlopt{+} \hldef{D} \hlopt{+} \hldef{E} \hlopt{~} \hldef{(A} \hlopt{+} \hldef{B} \hlopt{+} \hldef{C} \hlopt{+} \hldef{D} \hlopt{+} \hldef{E)}\hlopt{^}\hlnum{2} \hlopt{-} \hldef{(A} \hlopt{+} \hldef{B} \hlopt{+} \hldef{C} \hlopt{+} \hldef{D} \hlopt{+} \hldef{E)}
\hldef{+ }  \hldef{)}
\hldef{+ }\hldef{)}
\hldef{> }\hldef{design}\hlopt{$}\hlkwd{round}\hldef{(}\hlkwc{m} \hldef{=} \hlnum{12}\hldef{,} \hlkwc{method} \hldef{=} \hlsng{"orthogonal"}\hldef{)}
\end{alltt}
\end{kframe}
\end{knitrout}

\subsubsection{OMARS designs}

\begin{figure}[ht]
\center
\includegraphics[width=1\linewidth]{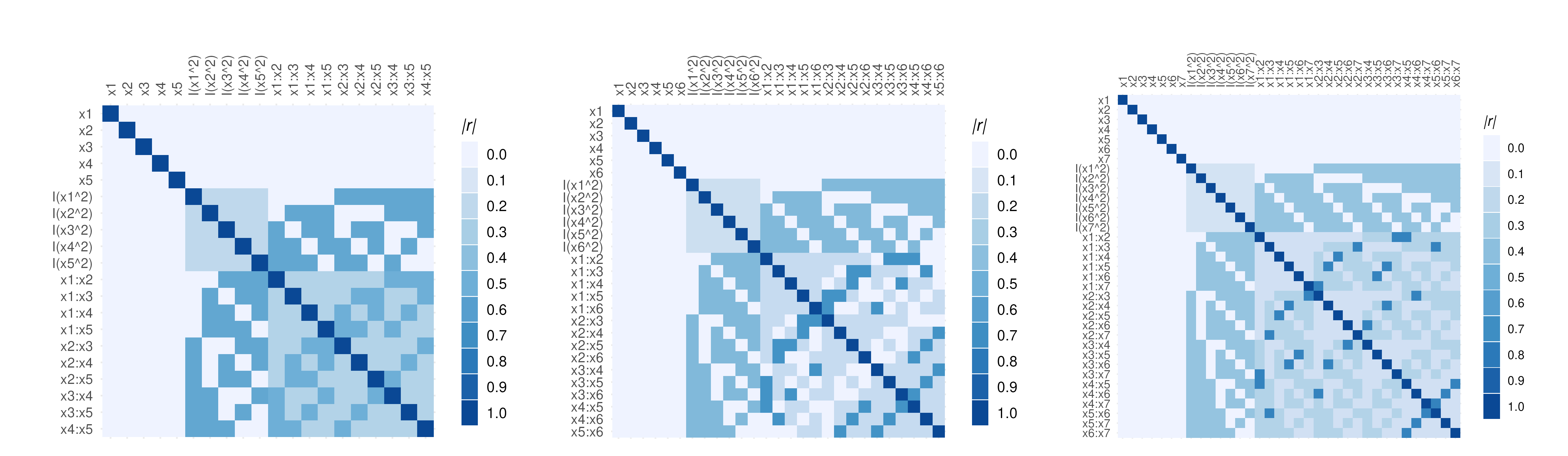}
\caption{Correlation maps of selected OMARS designs. From left to right: 5 factors ($m=12$); 6 factors ($m=16$); 7 factors ($m=20$).} \label{fig:4}
\end{figure}

Another class of modern designs that can be generated using this paradigm are the Orthogonal Minimally Aliased Response Surface (OMARS) designs introduced in \cite{NG20}. A standard OMARS design assumes a set of 3-level factors encoded as $\{-1, 0, +1\}$ to construct the candidate design grid and a second-order (full quadratic) linear model. The main effects columns in an OMARS design are constrained to be orthogonal to all other design columns, the number of zeros in all main effects columns are equal ($n_0^{ME}$) and the same for the number of zeros in all interaction columns ($n_0^{IE}$), and the average level of each factor is zero. Figure \ref{fig:4} shows the correlation maps of several OMARS designs obtained using the \code{orthogonal} rounding procedure for an increasing number of factors and number of runs, see Appendix \ref{app:B} for the relevant \proglang{R}-code.
The time to solve the MILP quickly explodes as the number of factors in the full quadratic model increases, as demonstrated by the benchmark timings in Appendix \ref{app:B}. In our experience, for up to 6 or 7 factors, the runtime of the MILP solver is in the order of minutes on a modern laptop computer. For higher problem dimensions, the MILP can be exported as an MPS (Mathematical Programming System) file and solved using an MILP solver with improved parallelization, such as the open-source GPU-accelerated \code{cuOpt} solver \citep{cuopt}, or other commercial solvers.

\section{Supported model types} \label{sec:3}

To solve the approximate optimal design problem from Section \ref{sec:2.1}, the matrix $M$ can be the information matrix of a model other than a standard linear regression model, given that $M(\boldsymbol{w})$ remains linear in the design weights $\boldsymbol{w}$. Several common model types are supported, each with a separate \pkg{R6}-class that is largely equivalent in its usage compared to the \code{lm_design} class for linear models. 

\subsection{Nonlinear and generalized linear models}

\subsubsection{Nonlinear models}

For nonlinear regression models with (i.i.d.) Gaussian errors, the \code{nlm_design} class is used to initialize a design object. An important difference compared to linear models, is that the information matrix of a nonlinear model $f$ is not independent of the model parameters $\boldsymbol{\theta} = (\theta_1, \ldots, \theta_p)$. If we denote $F(\boldsymbol{x}_i, \boldsymbol{\theta})$ for the gradient vector of the nonlinear model $f$ with respect to the model parameters $\boldsymbol{\theta}$ evaluated at $\boldsymbol{x}_i$, i.e. the $i$-th row of the Jacobian matrix. Then, conditional on the error variance $\sigma_e^2$, the $(p \times p)$-information matrix is:
\begin{eqnarray}
M(\boldsymbol{w}\,|\,\boldsymbol{X}, \boldsymbol{\theta}, \boldsymbol{v}) \ = \ \frac{1}{\sigma_e^2} \sum_{i = 1}^n w_i v_i F(\boldsymbol{x}_i, \boldsymbol{\theta}) F(\boldsymbol{x}_i, \boldsymbol{\theta})^T
\end{eqnarray}
which simplifies to the linear model information matrix if $f$ is linear in $\boldsymbol{\theta}$. In the general nonlinear case, only \emph{locally} optimal designs are calculated, conditional on a single vector of nominal parameter values $\boldsymbol{\theta}_0$. This is different from Bayesian optimal designs for nonlinear models, where an expected optimality criterion is calculated by integrating over a (prior) distribution assigned to the parameters $\boldsymbol{\theta}$.

\subsubsection{Generalized linear models}

For generalized linear and additive models with errors coming from an exponential dispersion family, the \code{glm_design} and \code{gam_design} classes should be used. Both classes are largely equivalent, except that the \code{gam_design} class uses \fct{mgcv::gam} to parse the design matrix associated to the additive model formula. For a generalized linear or additive model, the information matrix, conditional on the dispersion parameter of the assumed error distribution, is given by:
\begin{eqnarray}
M(\boldsymbol{w} \,|\, \boldsymbol{X}, \boldsymbol{\theta}, F, g, \boldsymbol{v}) \ = \ \sum_{i = 1}^n w_i v_i u(\boldsymbol{x}_i, \boldsymbol{\theta}, F, g) \boldsymbol{x}_i \boldsymbol{x}_i^T
\end{eqnarray}
The weights $u(\boldsymbol{x}_i, \boldsymbol{\theta}, F, g)$ capture the heteroscedasticity of the error distribution $F$ and depend on the link function $g$ and systematic (linear model) component $\eta(\boldsymbol{x}) = \boldsymbol{x}^T \boldsymbol{\theta}$, where $\boldsymbol{\theta}$ is the vector of model parameters. The exact formulation is given in Appendix \ref{app:A}. For a normal error distribution with identity link function, $u(\boldsymbol{x}_i, \boldsymbol{\theta}, F, g) = \frac{1}{\sigma_e^2}$ is constant and we recover the usual information matrix of a linear model. 
In general, for a heteroscedastic error distribution or non-identity link function, the information matrix depends on the model parameters $\boldsymbol{\theta}$ through $u(\boldsymbol{x}_i, \boldsymbol{\theta}, F, g)$. For this reason, only locally optimal designs are calculated, conditional on a vector of nominal parameter values $\boldsymbol{\theta}_0$, similar to the locally optimal designs for nonlinear models.

\subsubsection{Lubricant data example}

The calculation of a locally I-optimal design for a nonlinear regression model is demonstrated using the Lubricant dataset from \cite[Appendix 1, A1.8]{BW88}. This dataset describes the kinetic viscosity of a lubricant ($y$) as a function of temperature $x_1$ (°C) and pressure $x_2$ (atm) by a 9-parameter empirical nonlinear model:
\begin{eqnarray}
f(x_1, x_2, \boldsymbol{\theta}) \ = \ \frac{\theta_1}{\theta_2 + x_1} + \theta_3 x_2 + \theta_4 x_2^2 + \theta_5 x_2^3 + (\theta_6 x_2 + \theta_7 x_2^3) \exp\left( \frac{-x_1}{\theta_8 + \theta_9 x_2^2} \right)
\end{eqnarray} 
Nonlinear least squares parameter estimates were obtained in \cite{BW88} as $\boldsymbol{\theta}_0 = \{ 1054.541, 206.546, 1.46, -0.26, 0.023, 0.401, 0.035, 57.405, -0.477\}$. Given the nonlinear model formula, nominal parameter values $\boldsymbol{\theta}_0$ and a candidate design grid, locally optimal approximate or exact designs are calculated with \fct{\$optimize} and \fct{\$round} in the same way as for a linear model:
\begin{figure}[t]
\center
\includegraphics[width=1\linewidth]{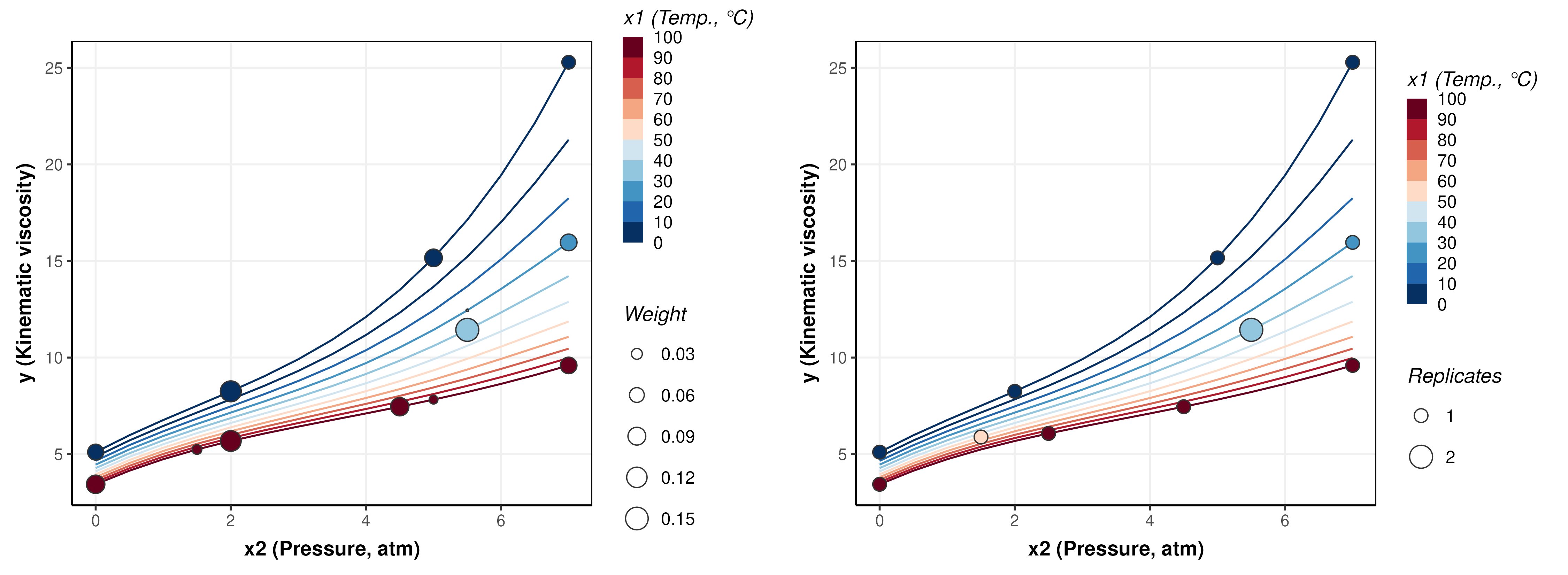}
\caption{Lubricant nonlinear model in the $(x_1, x_2, y)$ space. Left: locally I-optimal approximate design. Right: exact I-optimal design of size $m = 12$. The curves are the expected viscosity responses $f(x_1, x_2, \boldsymbol{\theta})$.} \label{fig:5}
\end{figure}
\begin{knitrout}
\definecolor{shadecolor}{rgb}{1, 1, 1}\color{fgcolor}\begin{kframe}
\begin{alltt}
\hldef{> }\hldef{design} \hlkwb{<-} \hldef{nlm_design}\hlopt{$}\hlkwd{new}\hldef{(}
\hldef{+ }  \hlkwc{formula} \hldef{=} \hlopt{~} \hldef{b1} \hlopt{/} \hldef{(b2} \hlopt{+} \hldef{x1)} \hlopt{+} \hldef{b3} \hlopt{*} \hldef{x2} \hlopt{+} \hldef{b4} \hlopt{*} \hldef{x2}\hlopt{^}\hlnum{2} \hlopt{+} \hldef{b5} \hlopt{*} \hldef{x2}\hlopt{^}\hlnum{3} \hlopt{+}
\hldef{+ }    \hldef{(b6} \hlopt{*} \hldef{x2} \hlopt{+} \hldef{b7} \hlopt{*} \hldef{x2}\hlopt{^}\hlnum{3}\hldef{)} \hlopt{*} \hlkwd{exp}\hldef{(}\hlopt{-}\hldef{x1} \hlopt{/} \hldef{(b8} \hlopt{+} \hldef{b9} \hlopt{*} \hldef{x2}\hlopt{^}\hlnum{2}\hldef{)),}
\hldef{+ }  \hlkwc{theta} \hldef{=} \hlkwd{c}\hldef{(}
\hldef{+ }    \hlkwc{b1} \hldef{=} \hlnum{1054.541}\hldef{,} \hlkwc{b2} \hldef{=} \hlnum{206.546}\hldef{,} \hlkwc{b3} \hldef{=} \hlnum{1.46}\hldef{,} \hlkwc{b4} \hldef{=} \hlopt{-}\hlnum{0.26}\hldef{,} \hlkwc{b5} \hldef{=} \hlnum{0.023}\hldef{,}
\hldef{+ }    \hlkwc{b6} \hldef{=} \hlnum{0.401}\hldef{,} \hlkwc{b7} \hldef{=} \hlnum{0.035}\hldef{,} \hlkwc{b8} \hldef{=} \hlnum{57.405}\hldef{,} \hlkwc{b9} \hldef{=} \hlopt{-}\hlnum{0.477}
\hldef{+ }  \hldef{),}
\hldef{+ }  \hlkwc{data} \hldef{=} \hlkwd{expand.grid}\hldef{(}
\hldef{+ }    \hlkwc{x1} \hldef{=} \hlkwd{seq}\hldef{(}\hlkwc{from} \hldef{=} \hlnum{0}\hldef{,} \hlkwc{to} \hldef{=} \hlnum{100}\hldef{,} \hlkwc{by} \hldef{=} \hlnum{10}\hldef{),}
\hldef{+ }    \hlkwc{x2} \hldef{=} \hlkwd{seq}\hldef{(}\hlkwc{from} \hldef{=} \hlnum{0}\hldef{,} \hlkwc{to} \hldef{=} \hlnum{7}\hldef{,} \hlkwc{by} \hldef{=} \hlnum{0.5}\hldef{)}
\hldef{+ }  \hldef{)}
\hldef{+ }\hldef{)}
\hldef{> }\hldef{design}\hlopt{$}\hlkwd{optimize}\hldef{(}\hlkwc{criterion} \hldef{=} \hlsng{"I"}\hldef{)}
\hldef{> }\hldef{design}\hlopt{$}\hlkwd{round}\hldef{(}\hlkwc{m} \hldef{=} \hlnum{12}\hldef{,} \hlkwc{method} \hldef{=} \hlsng{"optimal"}\hldef{)}
\end{alltt}
\end{kframe}
\end{knitrout}
The other methods and method arguments described in Section \ref{sec:2} can also be used for nonlinear models, except for the \code{orthogonal} rounding procedure, which is only available in a linear model context. 

An example of a designed experiment from \cite{LMM01} to count the number of defects in the manufacturing of semiconductor wafers is included in Appendix \ref{app:C}. A locally D-optimal design is generated using the \code{glm_design} class for a generalized linear model with a Poisson distribution and log-link function. 

\subsection{Random intercept linear mixed models}

Another important family of parametric models in industrial applications are Gaussian linear mixed models of the form $\boldsymbol{X}\boldsymbol{\theta} + \boldsymbol{Z}\boldsymbol{u} + \boldsymbol{\epsilon}$. The random effect covariates in $\boldsymbol{Z}$ represent nuisance factors that are not of primary interest or uncontrollable by the experimenter, but still influence the response by introducing correlation between observations with the same factor levels (e.g. batch or subject effects, or varying environmental conditions). If the nuisance factors in $\boldsymbol{Z}$ are ignored in the design, the estimation error in the model parameters of interest $\boldsymbol{\theta}$ may be sub-optimal due to the random effects correlation between observations. The \code{lme_design} \pkg{R6}-class supports design optimization only for the class of \emph{random intercept} linear mixed models with a single random intercept variance component $\sigma_u^2$. In this context, the covariance matrix of the observations $\text{Cov}(y)$ is block-diagonal of the form:
\begin{eqnarray}
\boldsymbol{V} \ = \ \sigma_e^2 \left(\boldsymbol{I}_n + \eta \boldsymbol{Z}\boldsymbol{Z}^T\right).
\end{eqnarray}
with $\sigma_e^2$ the variance of the (i.i.d.) errors $\boldsymbol{\epsilon}$, and the ratio $\eta = \sigma_u^2/\sigma_e^2$ representing the relative amount of correlation within random effects levels \citep[Chapter 3]{G02}.
Conditional on the variance ratio $\eta$, the $(p \times p)$-information matrix is given by:
\begin{eqnarray} \label{eq:15}
M(\boldsymbol{w}\,|\,\boldsymbol{X}, \boldsymbol{v}, \boldsymbol{Z}, \eta) \ = \frac{1}{\sigma_e^2} \left[ \sum_{i=1}^n w_i v_i \boldsymbol{x}_i \boldsymbol{x}_i^T - \sum_{j = 1}^b \frac{\eta}{1 + \eta k_j} \boldsymbol{S}_j \boldsymbol{S}_j^T  \right]
\end{eqnarray}
where $b$ is the number of blocks in $\boldsymbol{V}$; $k_j = \sum_{i \in B_j} w_iv_i$ is the sum of all weights in block $j$; and $\boldsymbol{S}_j = \sum_{i \in B_j} w_iv_i \boldsymbol{x}_i$ is the weighted sum of all design points in block $j$. If $\eta = 0$, the second term in the equation is zero and we recover the information matrix of a linear model. If $\eta > 0$, the information matrix is no longer linear in the design weights, which causes the usual formulation of eq.(\ref{eq:1}) as a convex optimization problem to become invalid. For this reason, the \fct{\$optimize} method of class \code{lme_design} implements a sequential convex approximation (SCA) procedure using a linearized version of the information matrix. The SCA procedure iterates through the following steps:
\begin{enumerate}
\item Initialize design weights $\boldsymbol{w}^{(0)}$ by solving the convex approximate design problem for $\eta = 0$.
\item Linearize the information matrix by a first-order Taylor approximation around $\boldsymbol{w}^{(0)}$ and solve the convex approximation for $\boldsymbol{w}^*$ by substituting the linearized information matrix, see Appendix \ref{app:A} for the mathematical details.
\item Choose new design weights $\boldsymbol{w}^{(1)} = \boldsymbol{w}^{(0)} + \delta (\boldsymbol{w}^* - \boldsymbol{w}^{(0)})$ using a backtracking line search to determine an appropriate step size $\delta$.
\item Repeat steps 2. and 3. until convergence or reaching the maximum number of iterations.
\end{enumerate}
If the candidate design grid is completely balanced, for instance when all combinations of the fixed and random effects factor levels are included as candidate design points, the effect of $\eta$ on the approximate optimal design is generally limited. As an artificial counter-example, consider the highly unbalanced candidate grid shown in Table \ref{tab:1} with one fixed effect discrete \code{Treatment} factor and one random effect categorical \code{Block} factor.
\begin{table}
\centering
\begin{tabular}[h]{rr}
\toprule
\code{Treatment} & \code{Block} \\
\midrule
-1 & \code{B1}\\
-1 & \code{B1}\\
0 & \code{B2}\\
0 & \code{B2}\\
1 & \code{B3}\\
\addlinespace
1 & \code{B3}\\
-1 & \code{B4}\\
0 & \code{B4}\\
0 & \code{B5}\\
1 & \code{B5}\\
\addlinespace
-1 & \code{B6}\\
1 & \code{B6}\\
\bottomrule
\end{tabular}
\caption{Unbalanced candidate design grid with three treatments (\code{Treatment}) and six blocks (\code{Block}).} \label{tab:1}
\end{table}
For a quadratic model in the \code{Treatment} variable without random effects, a D-optimal design only requires that each of the levels $\{-1, 0, 1\}$ is assigned equal weight. In particular, an approximate design assigning equal weight to each design point is D-optimal for a quadratic model. However, if the \code{Block} factor is included as a random effect, a D-optimal design should move more weight to the heterogeneous blocks \code{B4}, \code{B5} and \code{B6} as the variance ratio $\eta$ increases, since the \code{Treatment} variable is perfectly correlated with the \code{Block} factor for the other blocks. The following \proglang{R}-code demonstrates the SCA procedure to generate a D-optimal approximate design with $\eta = 10$. The obtained criterion value can be compared to that of a linear model D-optimal design with uniform weights using the \fct{\$crit} method:
\begin{knitrout}
\definecolor{shadecolor}{rgb}{1, 1, 1}\color{fgcolor}\begin{kframe}
\begin{alltt}
\hldef{> }\hldef{design} \hlkwb{<-} \hldef{lme_design}\hlopt{$}\hlkwd{new}\hldef{(}
\hldef{+ }  \hlkwc{formula} \hldef{=} \hlopt{~} \hldef{treatment} \hlopt{+} \hlkwd{I}\hldef{(treatment}\hlopt{^}\hlnum{2}\hldef{)} \hlopt{~} \hldef{block,}
\hldef{+ }  \hlkwc{eta} \hldef{=} \hlnum{10}\hldef{,}
\hldef{+ }  \hlkwc{data} \hldef{=} \hlkwd{data.frame}\hldef{(}
\hldef{+ }    \hlkwc{treatment} \hldef{=} \hlkwd{c}\hldef{(}\hlopt{-}\hlnum{1}\hldef{,} \hlopt{-}\hlnum{1}\hldef{,} \hlnum{0}\hldef{,} \hlnum{0}\hldef{,} \hlnum{1}\hldef{,} \hlnum{1}\hldef{,} \hlopt{-}\hlnum{1}\hldef{,} \hlnum{0}\hldef{,} \hlnum{0}\hldef{,} \hlnum{1}\hldef{,} \hlopt{-}\hlnum{1}\hldef{,} \hlnum{1}\hldef{),}
\hldef{+ }    \hlkwc{block} \hldef{=} \hlkwd{factor}\hldef{(}\hlkwd{rep}\hldef{(}\hlkwd{c}\hldef{(}\hlsng{"B1"}\hldef{,} \hlsng{"B2"}\hldef{,} \hlsng{"B3"}\hldef{,} \hlsng{"B4"}\hldef{,} \hlsng{"B5"}\hldef{,} \hlsng{"B6"}\hldef{),} \hlkwc{each} \hldef{=} \hlnum{2}\hldef{))}
\hldef{+ }  \hldef{)}
\hldef{+ }\hldef{)}
\hldef{> }\hldef{design}\hlopt{$}\hlkwd{optimize}\hldef{(}\hlkwc{criterion} \hldef{=} \hlsng{"D"}\hldef{,} \hlkwc{max_iter} \hldef{=} \hlnum{10}\hldef{)}
\hldef{> }\hldef{design}\hlopt{$}\hlkwd{crit}\hldef{(}\hlkwd{rep}\hldef{(}\hlnum{1}\hldef{,} \hlkwc{times} \hldef{=} \hlnum{12}\hldef{),} \hlkwc{criterion} \hldef{=} \hlsng{"D"}\hldef{)}
\end{alltt}
\end{kframe}
\end{knitrout}
\subsubsection{Split-plot designs}
A common scenario is the presence of \emph{hard-to-change} variables in the data that can only vary between blocks in the nuisance factors $\boldsymbol{Z}$, (e.g. manufacturing sites can only vary between different batches). In this context, a split-plot design \citep{G02} should be calculated, where the random effects factors are the \emph{whole plot} variables and the blocks are the individual whole plots, such that the levels of the hard-to-change variables remain fixed within each whole plot. Mathematically, if $\boldsymbol{H}$ denotes the design matrix of the hard-to-change factors, then the split-plot design must satisfy:
\begin{eqnarray}
\boldsymbol{H} = \boldsymbol{Z} \boldsymbol{H}_b
\end{eqnarray}
where $\boldsymbol{H}_b$ is a matrix with $b$ rows selected from the levels of the hard-to-change variables and $b$ is the number whole plots. The \code{lme_design} class includes a dedicated \fct{\$round\_split\_plot} method for the calculation of exact split-plot designs, which is demonstrated using a design case study from \cite{OM25}. 

\begin{figure}[t]
\center
\includegraphics[width=0.9\linewidth]{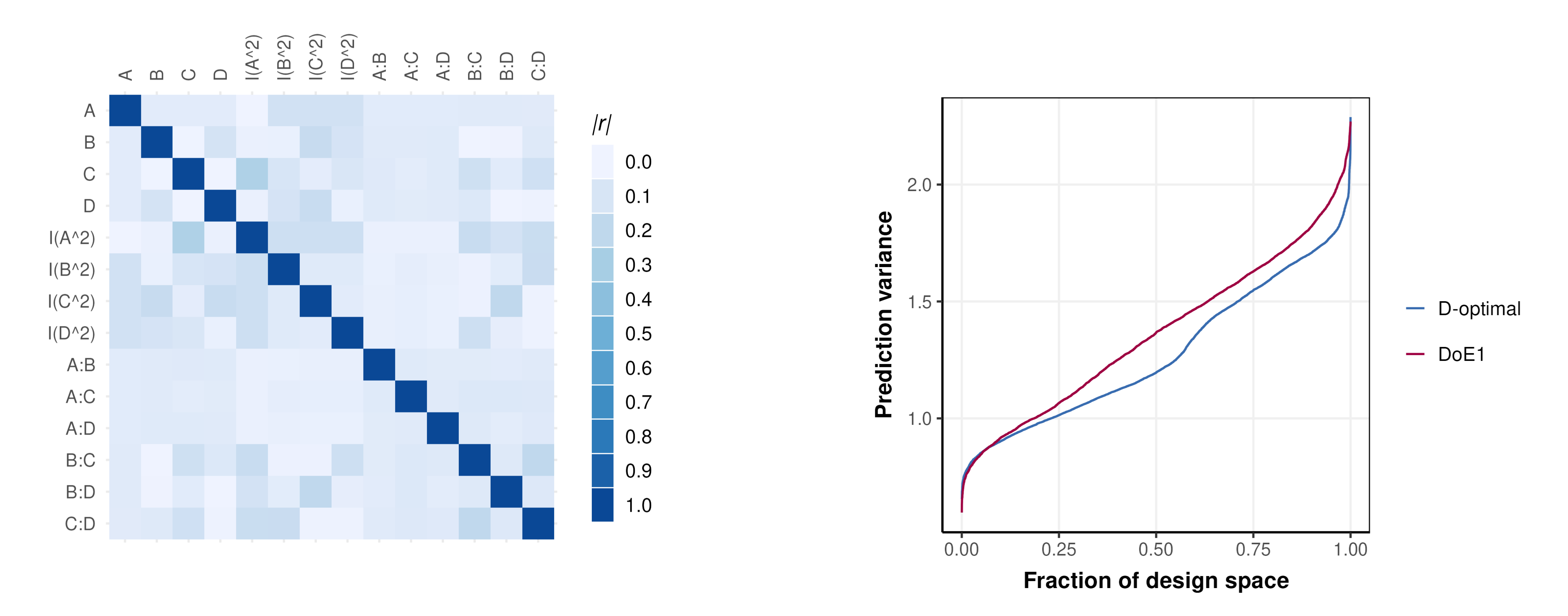}
\caption{Split-plot design diagnostic plots. Left: correlation map of 21-run D-optimal split-plot design. Right: fraction of design space plot.} \label{fig:6}
\end{figure}

The objective of the case study is to generate an experiment design for the synthesis of an active pharmaceutical ingredient. In total, there are 7 weeks available for the analysis, and a maximum of 3 experiments can be performed each week. There are four discrete factors of interest (\code{A}, \code{B}, \code{C}, and \code{D}), which are all encoded by 3 levels ($\{ -1, 0, 1\}$). A linear mixed model is assumed with a full quadratic model for the four fixed effects factors and a single random week intercept with variance component ratio $\eta = 1$ to capture the week-to-week variation:
\begin{knitrout}
\definecolor{shadecolor}{rgb}{1, 1, 1}\color{fgcolor}\begin{kframe}
\begin{alltt}
\hldef{> }\hldef{design} \hlkwb{<-} \hldef{lme_design}\hlopt{$}\hlkwd{new}\hldef{(}
\hldef{+ }  \hlkwc{formula} \hldef{=} \hlopt{~} \hldef{(A} \hlopt{+} \hldef{B} \hlopt{+} \hldef{C} \hlopt{+} \hldef{D)}\hlopt{^}\hlnum{2} \hlopt{+} \hlkwd{I}\hldef{(A}\hlopt{^}\hlnum{2}\hldef{)} \hlopt{+} \hlkwd{I}\hldef{(B}\hlopt{^}\hlnum{2}\hldef{)} \hlopt{+} \hlkwd{I}\hldef{(C}\hlopt{^}\hlnum{2}\hldef{)} \hlopt{+} \hlkwd{I}\hldef{(D}\hlopt{^}\hlnum{2}\hldef{)} \hlopt{~} \hldef{week,}
\hldef{+ }  \hlkwc{eta} \hldef{=} \hlnum{1}\hldef{,}
\hldef{+ }  \hlkwc{data} \hldef{=} \hlkwd{expand.grid}\hldef{(}
\hldef{+ }    \hlkwc{week} \hldef{=} \hlkwd{factor}\hldef{(}\hlnum{1}\hlopt{:}\hlnum{7}\hldef{,} \hlkwc{levels} \hldef{=} \hlnum{1}\hlopt{:}\hlnum{7}\hldef{),}
\hldef{+ }    \hlkwc{A} \hldef{=} \hlopt{-}\hlnum{1}\hlopt{:}\hlnum{1}\hldef{,}
\hldef{+ }    \hlkwc{B} \hldef{=} \hlopt{-}\hlnum{1}\hlopt{:}\hlnum{1}\hldef{,}
\hldef{+ }    \hlkwc{C} \hldef{=} \hlopt{-}\hlnum{1}\hlopt{:}\hlnum{1}\hldef{,}
\hldef{+ }    \hlkwc{D} \hldef{=} \hlopt{-}\hlnum{1}\hlopt{:}\hlnum{1}
\hldef{+ }  \hldef{)}
\hldef{+ }\hldef{)}
\end{alltt}
\end{kframe}
\end{knitrout}
The experiments are prepared on a weekly basis and operational constraints make it infeasible to work with multiple levels of factor \code{A} within a single week. A D-optimal split-plot design with 3 runs per week and a single hard-to-change factor \code{A} is generated as follows:
\begin{knitrout}
\definecolor{shadecolor}{rgb}{1, 1, 1}\color{fgcolor}\begin{kframe}
\begin{alltt}
\hldef{> }\hldef{design}\hlopt{$}\hlkwd{optimize}\hldef{(}\hlkwc{criterion} \hldef{=} \hlsng{"D"}\hldef{,} \hlkwc{max_iter} \hldef{=} \hlnum{5}\hldef{,} \hlkwc{return_value} \hldef{=} \hlnum{FALSE}\hldef{)}
\hldef{> }\hldef{design}\hlopt{$}\hlkwd{round_split_plot}\hldef{(}
\hldef{+ }  \hlkwc{m} \hldef{=} \hlnum{3}\hldef{,}
\hldef{+ }  \hlkwc{hard_to_change} \hldef{=} \hlopt{~}\hldef{A,}
\hldef{+ }  \hlkwc{n_repeats} \hldef{=} \hlnum{100}
\hldef{+ }\hldef{)}
\end{alltt}
\end{kframe}
\end{knitrout}
Compared to the initial split-plot design (DoE1) presented in \cite{OM25}, the D-optimal split-plot design has an increased D-efficiency of 8.7\% measured by the ratio of D-criteria and an increased I-efficiency of 21.3\% in terms of the ratio of I-criteria. Note that the initial split-plot design in \cite{OM25} was optimized using a composite D-optimality criterion \citep{MGG20}, so it is also not expected to be either D- or I-optimal in the ordinary sense. Figure \ref{fig:6} displays the correlation map of the D-optimal split-plot design, as well as a \emph{fraction of design space} plot visualizing the empirical cumulative distribution function of the prediction variance over the design space \citep[Chapter 4]{GJ11}, where the square roots of the prediction variances are calculated with the \fct{\$sep} method. The initial split-plot design is extended in \cite{OM25} to include one additional week with only 2 experiment runs. This scenario is handled by \fct{\$round\_split\_plot} by passing a vector \code{m = c(3, 3, 3, 3, 3, 3, 3, 2)} containing the individual whole plot sizes assigned to each week.

\section{Optimal design augmentation} \label{sec:4}

The implicit assumption in the design optimization problem of Section \ref{sec:2} is that the design is generated before conducting the experiments, without knowledge of the experimental response. In practice, experimentation may take place in multiple consecutive steps, in which case it is desirable to take into account the information present in an initial design $\boldsymbol{w}_0$ when constructing a new augmented optimal design $\boldsymbol{w}^*$. In the context of design augmentation, the approximate optimal design problem of Section \ref{sec:2.1} is updated to:
\begin{eqnarray} \label{eq:20}
\boldsymbol{w}^* \ = \ \arg \max_{\boldsymbol{w} \in \Delta^n} \Phi(M(\boldsymbol{w}\,|\, \boldsymbol{w}_0) + \lambda I) - \alpha \cdot \boldsymbol{c}^T\boldsymbol{w}
\end{eqnarray}
where the conditional information matrix $M(\boldsymbol{w}\,|\, \boldsymbol{w}_0) = M(\boldsymbol{w}\,|\, \boldsymbol{w}_0, \boldsymbol{X},\boldsymbol{v}, \gamma)$ is defined as a weighted linear combination of the two individual information matrices \citep[Chapter 11]{P06}. For a linear model, the conditional information matrix takes the form:
\begin{eqnarray}
M(\boldsymbol{w}\,|\, \boldsymbol{w}_0) \ = \ (1 - \gamma) M(\boldsymbol{w}_0) + \gamma M(\boldsymbol{w}) \ = \ \frac{1}{\sigma_e^2} \sum_{i=1}^n \tilde{w}_i v_i \boldsymbol{x}_i \boldsymbol{x}_i^T,
\end{eqnarray}
where $\tilde{w}_i = (1-\gamma) w_{i, 0} + \gamma w_i$. For an exact design, the weight contribution $\gamma$ is set to $\tfrac{m_1}{m_0 + m_1}$ in \cite[Chapter 11]{P06}, with $m_0$ the size of the initial design and $m_1$ the size of the new design. For an approximate design, $\gamma$ can be set to 0.5 equally weighting the old and new design, or $m_1$ can be set to the \emph{intended} size of the new design. Since the weights $\boldsymbol{w}_0$ are given, the augmented design optimization problem can be solved using the same SOCP formulation as for the original problem by modifying the lower bound constraint on the design weights, see Appendix \ref{app:A} for additional details. 

\subsubsection{Bayesian perspective}

A heuristic explanation for $M(\boldsymbol{w}\,|\,\boldsymbol{w}_0)$ being a linear combination of the information matrices of $\boldsymbol{w}_0$ and $\boldsymbol{w}$ is from the viewpoint of a Bayesian linear regression model. Let $\boldsymbol{X}$ be the design matrix associated to an exact design $\boldsymbol{\xi}$ of size $m$, and assume a standard linear regression model:
\begin{eqnarray}
\boldsymbol{y}\ \sim \ \boldsymbol{X}\boldsymbol{\theta} + \boldsymbol{\epsilon}, \quad \boldsymbol{\epsilon} \sim N(\boldsymbol{0}, \sigma_e^2 I_m), 
\end{eqnarray} 
with the parameter vector $\boldsymbol{\theta}$ having a normal prior $N(\boldsymbol{\theta}_0, \boldsymbol{\Sigma}_0)$, encoding all prior knowledge after observing the initial design $\boldsymbol{\xi}_0$. The posterior covariance matrix of $\boldsymbol{\theta}|\boldsymbol{y}$, conditional on the error variance $\sigma_e^2$, has the following analytic form:
\begin{eqnarray} \label{eq:23}
\boldsymbol{\Sigma}\, |\, \sigma_e, \boldsymbol{\Sigma}_0 \ = \ \left( \boldsymbol{\Sigma}_0^{-1} + \frac{1}{\sigma_e^2} \boldsymbol{X}^T \boldsymbol{X} \right)^{-1},
\end{eqnarray}
The conditional information matrix coincides with the inverse of the posterior covariance matrix. Therefore, using that $M(\boldsymbol{\xi}) = \frac{1}{\sigma_e^2} \boldsymbol{X}^T \boldsymbol{X}$ and $M(\boldsymbol{\xi}_0) = \boldsymbol{\Sigma}_0^{-1}$, it follows that $M(\boldsymbol{\xi}\,|\,\boldsymbol{\xi}_0)$ is exactly the sum of the individual information matrices.

\subsubsection{Active learning}

For the Bayesian linear regression model, a closed-form expression is available for the expected information gain $I(\boldsymbol{\xi}\,|\,\boldsymbol{\xi}_0)$, conditional on $\sigma_e$ and $\boldsymbol{\Sigma}_0$. Maximizing the expected information gain is a common information-theoretic query strategy in the context of \emph{active learning} \citep{S09}, which refers to online design augmentation based on a pre-defined utility function. The expected information gain $I(\boldsymbol{\xi}\,|\,\boldsymbol{\xi}_0)$ is defined as the Kullback-Leibler divergence between the prior and posterior distributions, which for normal distributions simplifies to:
\begin{eqnarray}
I(\boldsymbol{\xi}\,|\, \boldsymbol{\xi}_0) & = & \frac{1}{2} \log(\det(\boldsymbol{\Sigma}_0)) + \frac{1}{2} \log(\det(\boldsymbol{\Sigma}^{-1}\,|\, \sigma_e, \boldsymbol{\Sigma}_0 )) \\
& = & \frac{1}{2} \log(\det(\boldsymbol{\Sigma}_0)) + \frac{p}{2} \log(\det(M(\boldsymbol{\xi\,|\, \boldsymbol{\xi}_0}))^{1/p})
\end{eqnarray}
The first term on the right-hand side is constant, the second term is identical to the (log)D-criterion of the conditional information matrix. Maximizing the expected information gain for a Bayesian linear regression model with a normal prior is therefore equivalent to calculating a D-optimal augmented design.

\subsubsection{Extraction experiment example}

Each of the supported model classes include an \fct{\$augment} method for approximate optimal design augmentation. As an example of design augmentation for a linear model, we revisit the augmented extraction experiment from \cite[Chapter 3]{GJ11}. In the experiment, the extraction yield of a lipopeptide from a microbial culture is measured by varying 6 culture parameters: presence of methanol solvent ($x_1$), presence of ethanol solvent ($x_2$), presence of propanol solvent ($x_3$), presence of butanol solvent ($x_4$), culture pH ($x_5$), and time in solution ($x_6$). All factor levels are encoded as $-1$ and $+1$ referring to the absence or presence of each solvent and a low or high level of the pH and time in solution respectively. Following \cite{GJ11}, a linear model including all main effects terms and a subset of 6 first-order interaction terms is assumed to model the extraction yield:
\begin{knitrout}
\definecolor{shadecolor}{rgb}{1, 1, 1}\color{fgcolor}\begin{kframe}
\begin{alltt}
\hldef{> }\hldef{candidates} \hlkwb{<-} \hlkwd{expand.grid}\hldef{(}
\hldef{+ }  \hlkwc{x1} \hldef{=} \hlkwd{c}\hldef{(}\hlopt{-}\hlnum{1}\hldef{,} \hlnum{1}\hldef{),} \hlkwc{x2} \hldef{=} \hlkwd{c}\hldef{(}\hlopt{-}\hlnum{1}\hldef{,} \hlnum{1}\hldef{),} \hlkwc{x3} \hldef{=} \hlkwd{c}\hldef{(}\hlopt{-}\hlnum{1}\hldef{,} \hlnum{1}\hldef{),}
\hldef{+ }  \hlkwc{x4} \hldef{=} \hlkwd{c}\hldef{(}\hlopt{-}\hlnum{1}\hldef{,} \hlnum{1}\hldef{),} \hlkwc{x5} \hldef{=} \hlkwd{c}\hldef{(}\hlopt{-}\hlnum{1}\hldef{,} \hlnum{1}\hldef{),} \hlkwc{x6} \hldef{=} \hlkwd{c}\hldef{(}\hlopt{-}\hlnum{1}\hldef{,} \hlnum{1}\hldef{)}
\hldef{+ }\hldef{)}
\hldef{> }\hldef{design} \hlkwb{<-} \hldef{lm_design}\hlopt{$}\hlkwd{new}\hldef{(}
\hldef{+ }  \hlkwc{formula} \hldef{=} \hlopt{~} \hldef{x1} \hlopt{+} \hldef{x2} \hlopt{+} \hldef{x3} \hlopt{+} \hldef{x4} \hlopt{+} \hldef{x5} \hlopt{+} \hldef{x6} \hlopt{+}
\hldef{+ }    \hldef{x2}\hlopt{:}\hldef{x3} \hlopt{+} \hldef{x1}\hlopt{:}\hldef{x6} \hlopt{+} \hldef{x2}\hlopt{:}\hldef{x6} \hlopt{+} \hldef{x3}\hlopt{:}\hldef{x6} \hlopt{+} \hldef{x4}\hlopt{:}\hldef{x6} \hlopt{+} \hldef{x1}\hlopt{:}\hldef{x4,}
\hldef{+ }  \hlkwc{data} \hldef{= candidates}
\hldef{+ }\hldef{)}
\end{alltt}
\end{kframe}
\end{knitrout}
Given the linear model and the candidate design grid, an initial design of size $m_0 = 12$ with orthogonal main effects design columns, as well as a design augmentation of size $m_1 = 8$ is proposed in \cite[Chapter 3]{GJ11}, both of which are included in Appendix \ref{app:D}. Using the same initial design of size $m_0 = 12$, an approximate D-optimal augmented design is calculated with the \fct{\$augment} method setting $\gamma = \tfrac{m_1}{m_0 + m_1} = 0.4$ as follows:
\begin{knitrout}
\definecolor{shadecolor}{rgb}{1, 1, 1}\color{fgcolor}\begin{kframe}
\begin{alltt}
\hldef{> }\hldef{init_design} \hlkwb{<-} \hlkwd{c}\hldef{(}
\hldef{+ }  \hlnum{0}\hldef{,} \hlnum{0}\hldef{,} \hlnum{0}\hldef{,} \hlnum{0}\hldef{,} \hlnum{0}\hldef{,} \hlnum{0}\hldef{,} \hlnum{0}\hldef{,} \hlnum{1}\hldef{,} \hlnum{1}\hldef{,} \hlnum{0}\hldef{,} \hlnum{0}\hldef{,} \hlnum{0}\hldef{,} \hlnum{0}\hldef{,} \hlnum{0}\hldef{,} \hlnum{1}\hldef{,} \hlnum{0}\hldef{,} \hlnum{0}\hldef{,} \hlnum{1}\hldef{,} \hlnum{1}\hldef{,} \hlnum{0}\hldef{,} \hlnum{0}\hldef{,} \hlnum{0}\hldef{,} \hlnum{0}\hldef{,} \hlnum{0}\hldef{,}
\hldef{+ }  \hlnum{0}\hldef{,} \hlnum{0}\hldef{,} \hlnum{0}\hldef{,} \hlnum{0}\hldef{,} \hlnum{0}\hldef{,} \hlnum{1}\hldef{,} \hlnum{0}\hldef{,} \hlnum{0}\hldef{,} \hlnum{0}\hldef{,} \hlnum{0}\hldef{,} \hlnum{0}\hldef{,} \hlnum{1}\hldef{,} \hlnum{1}\hldef{,} \hlnum{0}\hldef{,} \hlnum{0}\hldef{,} \hlnum{0}\hldef{,} \hlnum{0}\hldef{,} \hlnum{1}\hldef{,} \hlnum{0}\hldef{,} \hlnum{0}\hldef{,} \hlnum{0}\hldef{,} \hlnum{0}\hldef{,} \hlnum{0}\hldef{,} \hlnum{0}\hldef{,}
\hldef{+ }  \hlnum{0}\hldef{,} \hlnum{0}\hldef{,} \hlnum{0}\hldef{,} \hlnum{0}\hldef{,} \hlnum{1}\hldef{,} \hlnum{0}\hldef{,} \hlnum{0}\hldef{,} \hlnum{0}\hldef{,} \hlnum{0}\hldef{,} \hlnum{0}\hldef{,} \hlnum{1}\hldef{,} \hlnum{0}\hldef{,} \hlnum{0}\hldef{,} \hlnum{0}\hldef{,} \hlnum{0}\hldef{,} \hlnum{1}
\hldef{+ }\hldef{)}
\hldef{> }\hldef{design}\hlopt{$}\hlkwd{augment}\hldef{(}
\hldef{+ }  \hlkwc{design_weights} \hldef{= init_design,}
\hldef{+ }  \hlkwc{gamma} \hldef{=} \hlnum{0.4}\hldef{,}
\hldef{+ }  \hlkwc{criterion} \hldef{=} \hlsng{"D"}
\hldef{+ }\hldef{)}
\end{alltt}
\end{kframe}
\end{knitrout}
For the calculation of an exact D-optimal augmented design, the \fct{\$round} method is used by passing the initial design vector to the \code{augment_design} argument. The \code{optimal} rounding method executes the same exchange algorithms as before, but excluding any of the design points in the initial design from being exchanged with new design points:
\begin{knitrout}
\definecolor{shadecolor}{rgb}{1, 1, 1}\color{fgcolor}\begin{kframe}
\begin{alltt}
\hldef{> }\hldef{design}\hlopt{$}\hlkwd{round}\hldef{(}
\hldef{+ }  \hlkwc{augment_design} \hldef{= init_design,}
\hldef{+ }  \hlkwc{m} \hldef{=} \hlnum{8}\hldef{,}
\hldef{+ }  \hlkwc{method} \hldef{=} \hlsng{"optimal"}
\hldef{+ }\hldef{)}
\end{alltt}
\end{kframe}
\end{knitrout}
Table \ref{tab:2} shows the exact design augmentation of size $m_1 = 8$ obtained with the above \proglang{R}-code. The D-optimal augmented design is nearly identical to the proposed design in \cite[Chapter 3]{GJ11}, except for one design point, the reason being that the D-criterion can be marginally improved (0.14\%) by substituting the point $(+1, -1, -1, -1, -1, -1)$.
\begin{table}[t]
\centering
\begin{tabular}{lrrrrrr}
\toprule
  & \code{x1} & \code{x2} & \code{x3} & \code{x4} & \code{x5} & \code{x6}\\
\midrule
13 & -1 & -1 & 1 & -1 & -1 & -1\\
14 & 1 & 1 & -1 & 1 & -1 & -1\\
15 & -1 & 1 & 1 & 1 & 1 & -1\\
16 & -1 & 1 & 1 & -1 & -1 & 1\\
17 & -1 & -1 & 1 & 1 & -1 & 1\\
19 & 1 & -1 & 1 & -1 & 1 & 1\\
18 & -1 & -1 & -1 & -1 & 1 & 1\\
20 & -1 & -1 & -1 & 1 & 1 & 1\\
\bottomrule
\end{tabular}
\caption{Extraction experiment, D-optimal design augmentation of size 8.} \label{tab:2}
\end{table}

\subsection{Penalized design augmentation} \label{sec:4.1}

A primary application of (sequential) penalized design augmentation is global optimization of an unknown response, where the cost penalties $\boldsymbol{c}$ or the weights $\boldsymbol{v}$ are informed by the current knowledge of the response across the candidate design grid. If the objective is response maximization, candidate points with a high expected response are favored when augmenting the design, resulting in a trade-off between exploration of the design space through the design criterion and exploitation of the response through the cost penalties or weights. The exploration versus exploitation paradigm is common in the context of Bayesian Optimization (BO), which can also be understood as a data-driven sequential design augmentation procedure. 

To make this more precise, consider a D-optimal design for a linear model with cost penalties $c_i = -\mu(\boldsymbol{x_i})$, where $\mu(\boldsymbol{x_i})$ is the expected response at $\boldsymbol{x}_i$ given the current data or prior knowledge. Then it is verified in Appendix \ref{app:A} that, at the optimal design $\boldsymbol{w}^*$, the balance between the prediction variance and the expected mean response is the same everywhere across the support of the design, which can be viewed as a particular version of the General Equivalence Theorem \citep{KW60} with heterogeneous costs informed by the expected response. 

As a consequence, if the design is augmented with a single design point, i.e. if a single design point proposal is generated, the optimal design maximizes the acquisition function:
\begin{eqnarray} \label{eq:30}
\text{acq}(\boldsymbol{x}_i) \ = \ \mu(\boldsymbol{x}_i) + \tau \sigma^2(\boldsymbol{x_i} | \boldsymbol{w}^*)
\end{eqnarray}
with $\tau$ depending on the penalty scaling parameter $\alpha$ and the number of parameters $p$. This acquisition function is remarkably similar to the Upper Confidence Bound (UCB) acquisition function in Bayesian Optimization \citep{S12}, except that the exploration term in the UCB acquisition function is the predictive standard deviation $\sigma(\boldsymbol{x}_i)$ instead of the predictive variance of the surrogate model. Another minor difference is that the D-optimal exploration term $\sigma^2(\boldsymbol{x}_i | \boldsymbol{w}^*)$ includes the information of the augmented design point, whereas the UCB exploration term $\sigma(\boldsymbol{x}_i)$ is the predictive standard deviation before augmenting the design.

For a linear model D-optimal design with heterogeneous weights $v_i = \mu(\boldsymbol{x}_i)$ and without cost penalties, when augmenting the design by a single design point, the optimal design maximizes the acquisition function:
\begin{eqnarray}
\text{acq}(\boldsymbol{x}_i) \ = \ \mu(\boldsymbol{x}_i) \sigma^2(\boldsymbol{x_i} | \boldsymbol{w}^*)
\end{eqnarray}
This acquisition function induces a multiplicative, instead of an additive, trade-off between exploitation and exploration terms. As previously noted in Section \ref{sec:2.3}, an advantage of the multiplicative trade-off is that the augmented design is invariant under scaling of $\boldsymbol{v}$.

\subsubsection{Imidazoles reaction example}

To demonstrate optimization of a response through sequential penalized design augmentation, we revisit the direct arylation reaction example from Section \ref{sec:2.3}. In this example, we focus on optimization of the reaction yield across the 1728 candidate experiments, excluding the experiment reaction costs used in Section \ref{sec:2.3}.  
The initial design is a regularized I-optimal design of size $m_0 = 15$ for a linear model including all main effects and first-order interactions between the discrete temperature and concentration factors and the categorical base, solvent and ligand factors. Since the number of model parameters ($p = 130$) is larger than the initial design size, a small ridge penalty $\lambda = 10^{-2}$ is included to regularize the information matrix. The I-optimal design is augmented 6 times by 10 new design points, using heterogeneous weights $\boldsymbol{v}$ at the $k$-th iteration that promote high responses according to:
\begin{eqnarray}
\boldsymbol{v}^{(k)}_i = \hat{y}(\boldsymbol{x}_i\, |\, \boldsymbol{\xi}^{(k)} ), \quad \text{for } i = 1,\ldots,n
\end{eqnarray}
where $\hat{y}(\boldsymbol{x}_i | \boldsymbol{\xi}^{(k)})$ is the predicted response at $\boldsymbol{x}_i$ after evaluation of the design $\boldsymbol{\xi}^{(k)}$. The predictions $\hat{y}(\boldsymbol{x}_i)$ are made by a lasso regression model fitted to the available data at iteration $k$ with \pkg{glmnet} \citep{glmnet}, using leave-one-out cross-validation to tune the lasso penalty parameter. A lasso prediction model was chosen because the full linear model cannot be fitted to a design of size $m < p$, as well as the belief that the yield can be predicted by a sparse model using only a small fraction of the total number of coefficients $p$. 
\begin{figure}[t]
\center
\includegraphics[width=1\linewidth]{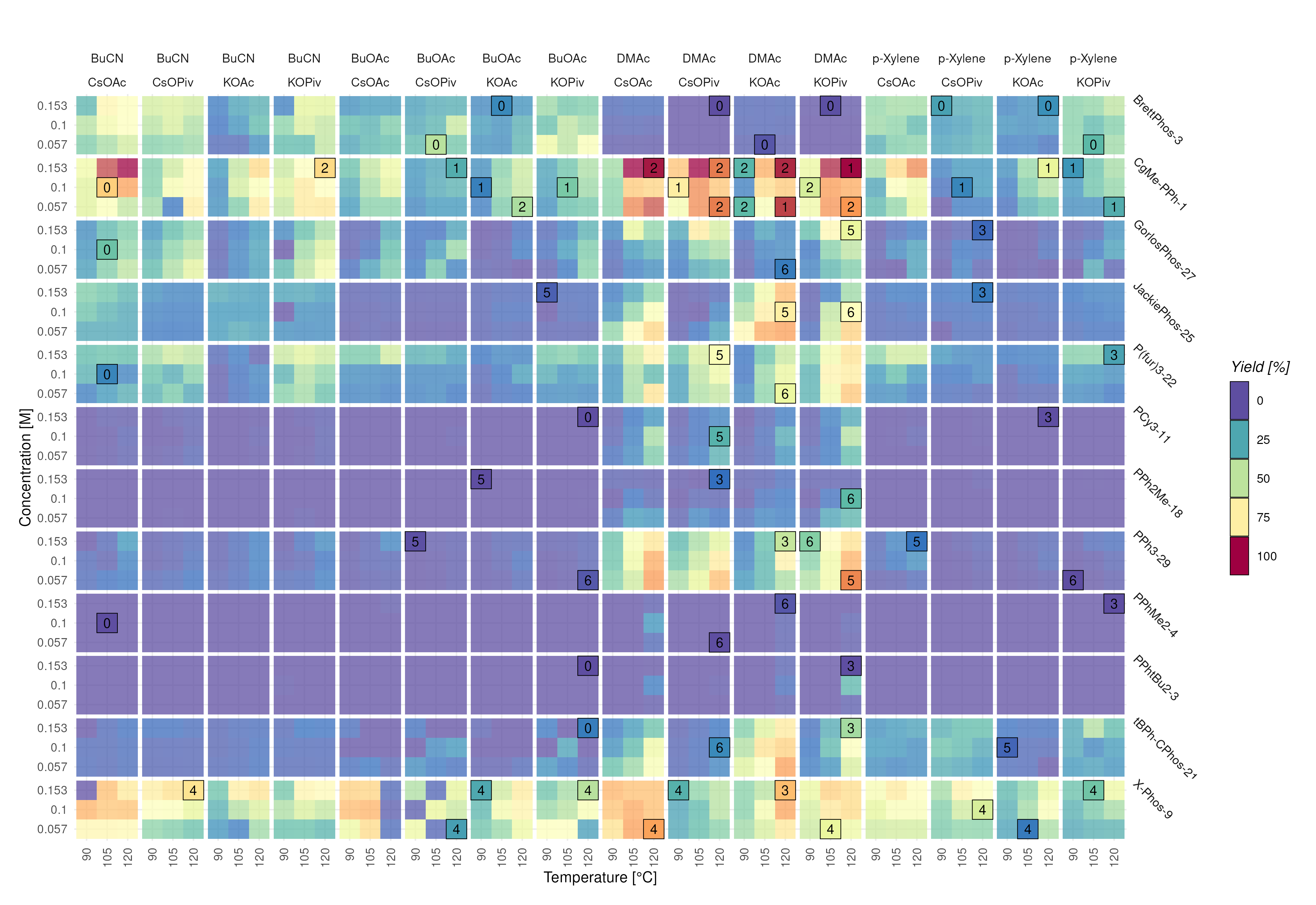}
\caption{Imidazoles reaction yields overlayed by point proposals from sequential weighted I-optimal augmentation (6 times 10 design points) of an initial I-optimal design of size 15.} \label{fig:8}
\end{figure}
Figure \ref{fig:8} displays the 1728 experimental yields overlayed with the locations of the proposed design points. The numbers inside the tiles refer to the iteration in which the design point was added to the augmented design, with zero referring to the initial regularized I-optimal design. The lasso regression model correctly identifies the CgMe-PPh-1 ligand and the DMAc solvent as being important predictors for a high yield, which results in a strong bias in the proposed design points towards these ligand and solvent levels. The maximum observed yield in the design is 99.81\% and the root mean squared error with respect to the target response surface is 17.28\% ($R^2 = 0.507$). Interestingly, the root mean squared error of the lasso regression given an unweighted regularized I-optimal design of size $m = 75$ is 17.06\% ($R^2 = 0.519$), which is only slightly lower. This is attributed to the fact that most of the signal is present in the high yield regions and exploration of the low yield regions only provides limited information to improve prediction accuracy. It should be noted that, similar to a Bayesian Optimization loop, there is no guarantee that the global maximum is found by the sequentially augmented design as evidenced by the fact that not all regions of the candidate design space with high yields are fully explored (e.g. the interaction between BuCN and CgMe-PPh-1 or X-Phos-9). Relevant \proglang{R}-code to replicate this example is included in Appendix \ref{app:C}.

\subsubsection{SARS-CoV-2 drug activity example}

A more advanced example of sequential design augmentation using a kernel-based classifier is presented using an imbalanced drug activity dataset \citep{TDC21} based on high-throughput screening data from \cite{D20}. The dataset reports binary levels of activity against SARS-CoV-2 3CL Protease for 880 drug compounds identified by their SMILES strings, consisting of 78 high- and 802 low-activity compounds. Viewing this as a drug discovery problem, our aim is to efficiently identify a maximum number of drug compounds with a high activity response, assuming no initial knowledge of the activity responses. 

To initialize the candidate design grid, binary (1024-bit) Morgan fingerprints were generated for each drug molecule from their SMILES strings using \pkg{RDKit} \citep{rdkit}. The initial design is a regularized D-optimal design for a linear model including only the fingerprint main effects ($p = 1024$) of size $m = 50$ using a small ridge penalty $\lambda = 10^{-2}$ to regularize the information matrix. After evaluation of the initial design, a kernel logistic regression model is fitted to the available data according to:
\begin{eqnarray}
&& y_i \ \sim \ \text{Bernoulli}(p_i), \quad \text{for } i = 1,\ldots,m \\
&& \text{with} \quad \text{logit}(p_i) \ = \  \phi(\boldsymbol{x}_i)^T \boldsymbol{\theta}
\end{eqnarray}
The latent coefficient vector $\boldsymbol{\theta}$ contains as many coefficients as the number of observations $m$, and $\phi(\boldsymbol{x}_i)$ is a kernel mapping of the binary fingerprint vector $\boldsymbol{x}_i$, such that $\phi(\boldsymbol{x}_i)^T \phi(\boldsymbol{x}_j) = k(\boldsymbol{x}_i, \boldsymbol{x}_j)$, with a parameterized Tanimoto (also Jaccard) kernel to measure similarity between binary fingerprints:
\begin{eqnarray}
k(\boldsymbol{x}_i, \boldsymbol{x}_j) \ = \  \sigma^2 \exp\left( -\frac{1 - T(\boldsymbol{x}_i, \boldsymbol{x}_j)}{\rho}\right), \quad \text{with} \quad T(\boldsymbol{x}_i, \boldsymbol{x}_j) \ = \ \frac{\boldsymbol{x}_i \cdot \boldsymbol{x}_j}{\Vert \boldsymbol{x}_i \Vert^2 + \Vert \boldsymbol{x}_j \Vert^2 - \boldsymbol{x}_i \cdot \boldsymbol{x}_j}
\end{eqnarray}
The kernel is parameterized by a magnitude $\sigma$ and a length-scale $\rho$, which are estimated together with $\boldsymbol{\theta}$ when fitting the logistic regression model. Given the estimated $(n \times m)$ kernel mapping matrix $\widehat{\Phi} = \big[ \hat{\phi}(\boldsymbol{x}_i),\ldots,\hat{\phi}(\boldsymbol{x}_n)\big]^T$ and the predicted probabilities $\hat{p}_1,\ldots,\hat{p}_n$ over the candidate design grid, a new \code{glm_design} object is initialized with design matrix $\widehat{\Phi}$ and nominal parameter values $\hat{\boldsymbol{\theta}}$ as: 
\begin{knitrout}
\definecolor{shadecolor}{rgb}{1, 1, 1}\color{fgcolor}\begin{kframe}
\begin{alltt}
\hldef{> }\hldef{design} \hlkwb{<-} \hldef{glm_design}\hlopt{$}\hlkwd{new}\hldef{(}
\hldef{+ }  \hlkwc{formula} \hldef{=} \hlopt{~} \hlnum{0} \hlopt{+} \hldef{.,}
\hldef{+ }  \hlkwc{data} \hldef{= Phi_hat,}
\hldef{+ }  \hlkwc{family} \hldef{=} \hlkwd{binomial}\hldef{(}\hlkwc{link} \hldef{=} \hlsng{"log"}\hldef{),}
\hldef{+ }  \hlkwc{theta} \hldef{= theta_hat,}
\hldef{+ }  \hlkwc{cost} \hldef{=} \hlnum{1} \hlopt{-} \hldef{p_hat}
\hldef{+ }\hldef{)}
\end{alltt}
\end{kframe}
\end{knitrout}
\begin{figure}[t]
\center
\includegraphics[width=1\linewidth]{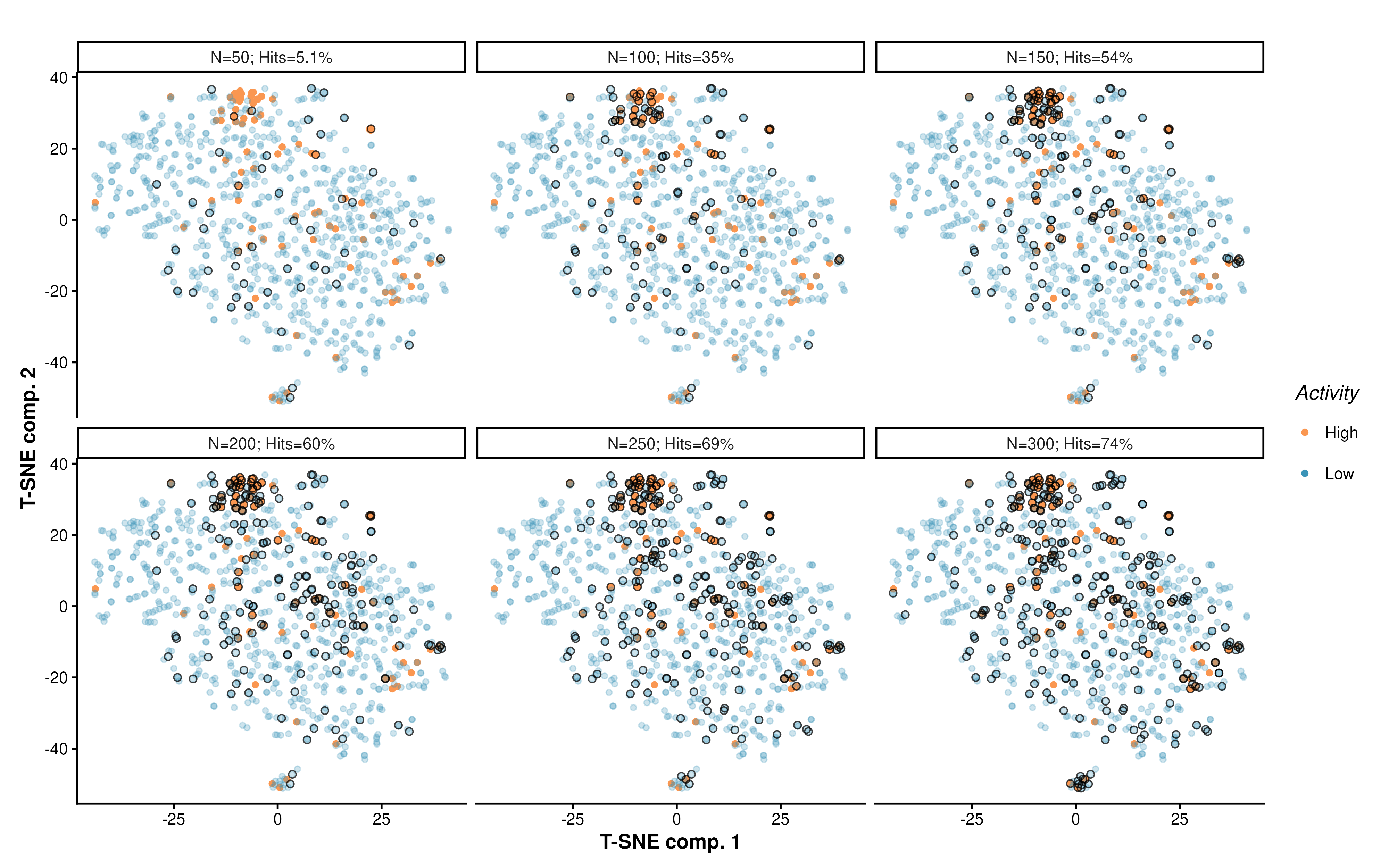}
\caption{Binary molecular fingerprints in 2D (T-SNE) space, highlighting proposals from sequential penalized D-optimal augmentation (5 times 50 compounds) using a kernel logistic model, with \code{Hits} the percentage of observed high-activity compounds.} \label{fig:9}
\end{figure}
The new design object imposes additive cost penalties $c(\boldsymbol{x}_i) = 1 - \hat{p}_i$, thereby penalizing compounds with a low predicted activity. The global penalty scaling parameter is kept at the default $\alpha = 1$. With the new design object, the existing design is augmented by $m_1 = 50$ new compounds using D-optimality as the optimization criterion. Figure \ref{fig:9} displays the percentage of identified high-activity compounds by augmenting the design 5 times with 50 compounds, updating the \code{data}, \code{theta} and \code{cost} arguments in the \code{glm_design} object after each iteration. The 1024-bit molecular fingerprints are visualized in 2D using unsupervised T-distributed Stochastic Neighbor Embedding (T-SNE). The open circles highlight the compounds included in the sequentially augmented design, showing the strong bias towards the cluster of high-activity compounds. More generally, the performance of the demonstrated online design approach depends heavily on the accuracy of the model predictions. In this example, identifying the complete set of high-activity compounds becomes increasingly hard due to the inability of the model to correctly predict the response of the unobserved high-activity compounds using only the binary molecular fingerprints. The kernel logistic regression model was fitted using \proglang{Stan} \citep{stan} by optimizing the maximum a posteriori (MAP) estimates of the parameters. An extended description on the use of (approximate) Gaussian process models for penalized sequential design augmentation is included in Appendix \ref{app:C}.

\section{Conclusions and prospects} \label{sec:5}

This work introduces \pkg{cvxoptdes} as an open-source \proglang{R}-package for general model-based design optimization and augmentation. A primary contribution of \pkg{cvxoptdes} is to unify classical optimal experiment design and online response optimization within a single framework. Instead of treating these as separate problems, both are combined in the form of a penalized design optimization problem to explicitly trade off exploration of the design space and optimization of the unknown response. This constitutes a novel, interpretable alternative to active learning and Bayesian optimization in settings where a structured parametric (linear or nonlinear) model is preferable to a data-driven nonparametric surrogate model. A second key contribution is the coverage of model classes and design types: \pkg{cvxoptdes} supports approximate and exact design optimization for weighted or ridge linear, generalized linear and additive, nonlinear, random-intercept linear mixed models, as well as the calculation of split-plot, stratified, and OMARS designs among others. Together, these contributions consolidate two traditionally separate strands of methodology, optimal design generation and online response optimization, into one coherent \proglang{R}-package, reducing the overhead for practitioners of learning and switching between multiple software programs and packages.

The \pkg{cvxoptdes} \proglang{R}-package was implemented using an object-oriented architecture with the idea in mind that new design criteria, model classes, and class methods could be added in the future. The design criteria discussed in this work all aim to optimize model identifiability or precision for a single candidate model, and are not well suited to settings where the design objective is to discriminate between multiple candidate models. Criteria for model discrimination, such as T-optimality \citep{AF75} or the Box-Hill criterion \citep{BH67} in a sequential design setting, are not currently implemented but can be incorporated in a modular fashion. Similarly, new model classes or class methods, such as linear mixed models with multiple variance components and multi-level split-plot designs, can be included in the future. Note that conditional multi-level split-plot designs can already be generated by calculating a separate split-plot design and fixing the hard-to-change levels at each level of nesting similar to \cite{MK21}. Finally, design penalization functions other than a linear penalty could be integrated in the package, further developing the equivalence between penalized design augmentation procedures and common acquisition strategies in Bayesian optimization or other online optimization approaches, such as information-directed sampling \citep{RR14}.

The \proglang{R}-package source code of \pkg{cvxoptdes} is available on GitHub (\url{https://cran.r-project.org/package=cvxoptdes}) and the \proglang{R}-package will be made available on CRAN.

\section*{Acknowledgments}

The authors gratefully acknowledge Tatsiana Khamiakova for initiating and supporting this work, and Jan Dijkmans for the initial motivation from a chemical setting. We also thank Vahid Nassiri and Tobias Verbeke for their suggestions that helped to improve the presentation of this work.

%% -- Bibliography -------------------------------------------------------------
%% - References need to be provided in a .bib BibTeX database.
%% - All references should be made with \cite, \citet, \citep, \citealp etc.
%%   (and never hard-coded). See the FAQ for details.
%% - JSS-specific markup (\proglang, \pkg, \code) should be used in the .bib.
%% - Titles in the .bib should be in title case.
%% - DOIs should be included where available.

\bibliography{refs}

%% -- Appendix (if any) --------------------------------------------------------
%% - After the bibliography with page break.
%% - With proper section titles and _not_ just "Appendix".

\newpage

\begin{appendix}

\section{Mathematical details} \label{app:A}

\subsection{Generalized linear models}

Recall that a generalized linear model is defined by an error distribution family $F$, a link function $g$ and a systematic (linear model) component $\eta(\boldsymbol{x}) = \boldsymbol{x}^T \boldsymbol{\theta}$, where $\boldsymbol{x}$ is a $p$-dimensional design point and $\boldsymbol{\theta}$ is the vector of model parameters of the same dimension. The error distribution family $F$ is assumed to be an exponential dispersion family meaning that the probability density (or mass) function of $F$ can be written in standard canonical form as:
\begin{eqnarray}
f(y\,|\, \zeta,\phi) & = & \exp\left( \tfrac{y\zeta - b(\zeta)}{a(\phi)} + c(y, \phi) \right).
\end{eqnarray}
Here, $b(\zeta(\boldsymbol{x}))$ is the cumulant generating function, such that $\mu(\boldsymbol{x}) = E[Y | \boldsymbol{x}] = b'(\zeta(\boldsymbol{x}))$ and $V(\mu(\boldsymbol{x})) = b''(\zeta(\boldsymbol{x}))$, where the latter is the variance function expressed as a function of the mean. The systematic component $\eta$ is linked to the mean $\mu$ through the link function $g$:
\begin{eqnarray}
g(\mu(\boldsymbol{x})) \ = \ \eta(\boldsymbol{x}) \ = \ \boldsymbol{x}^T \boldsymbol{\theta}
\end{eqnarray}
Given the  $(n \times p)$ design matrix $\boldsymbol{X}$ with rows $\boldsymbol{x}_1,\ldots,\boldsymbol{x}_n$, the $(p \times p)$-dimensional information matrix, conditional on the dispersion parameter $\phi$, is given by:
\begin{eqnarray}
M(\boldsymbol{w} \,|\, \boldsymbol{X}, \boldsymbol{\theta}, F, g, \boldsymbol{v}) \ = \ \sum_{i = 1}^n w_i v_i u(\boldsymbol{x}_i, \boldsymbol{\theta}, F, g) \boldsymbol{x}_i \boldsymbol{x}_i^T
\end{eqnarray}
with weights $u(\boldsymbol{x}_i, \boldsymbol{\theta}, F, g)$ capturing the heteroscedasticity of the error distribution according to:
\begin{eqnarray}
u(\boldsymbol{x}_i, \boldsymbol{\theta}, F, g) \ = \ \frac{1}{a(\phi)} \frac{1}{V(\mu(\boldsymbol{x}_i))[g'(\mu(\boldsymbol{x}_i))]^2}, \quad \text{for } i = 1,\ldots,n
\end{eqnarray}
For a normal error distribution with identity link function, $V(\mu(\boldsymbol{x}_i)) = 1$ and $g'(\mu(\boldsymbol{x}_i)) = 1$, resulting in constant weights $u(\boldsymbol{x}_i, \boldsymbol{\theta}, F, g) = \frac{1}{a(\phi)} = \frac{1}{\sigma_e^2}$ equal to the inverse of the error variance. 

\subsection{Linear mixed models}

\subsubsection{Sequential convex approximation}

Recall that, conditional on the variance ratio $\eta = \sigma_u^2 / \sigma_e^2$, the $(p \times p)$-information matrix is:
\begin{eqnarray} \label{app_eq:1}
M(\boldsymbol{w}\,|\,\boldsymbol{X}, \boldsymbol{v}, \boldsymbol{Z}, \eta) \ = \frac{1}{\sigma_e^2} \left[ \sum_{i=1}^n w_i v_i \boldsymbol{x}_i \boldsymbol{x}_i^T - \sum_{j = 1}^b \frac{\eta}{1 + \eta k_j} \boldsymbol{S}_j \boldsymbol{S}_j^T  \right]
\end{eqnarray}
where $b$ is the number of blocks in $\boldsymbol{V}$; $k_j = \sum_{i \in B_j} w_iv_i$ is the sum of all weights in block $j$; and $\boldsymbol{S}_j = \sum_{i \in B_j} w_iv_i \boldsymbol{x}_i$ is the weighted sum of all design points in block $j$. 

In order to calculate the partial derivative of the information matrix with respect to $w_i$, with $i \in B_j$, we derive both terms in eq.(\ref{app_eq:1}). For the first term,
\begin{eqnarray*}
\frac{\partial}{\partial w_i}\left(\sum_{i = 1}^n w_iv_i \boldsymbol{x}_i\boldsymbol{x}_i^T \right) &=&  v_i\boldsymbol{x}_i\boldsymbol{x}_i^T 
\end{eqnarray*}
For the second term, denoting $\zeta_j = \frac{\eta}{1 + \eta k_j}$,
\begin{eqnarray*}
\frac{\partial}{\partial w_i}\left(\sum_{j = 1}^b \zeta_j \boldsymbol{S}_j \boldsymbol{S}_j^T \right) &=& \frac{\partial \zeta_j }{\partial w_i} \boldsymbol{S}_j \boldsymbol{S}_j^T + \zeta_j \frac{\partial \left( \boldsymbol{S}_j \boldsymbol{S}_j^T \right) }{\partial w_i} \\
&=& -v_i \zeta_j^2 \boldsymbol{S}_j \boldsymbol{S}_j^T + v_i \zeta_j \left[ \boldsymbol{x}_i \boldsymbol{S}_j^T + \boldsymbol{S}_j \boldsymbol{x}_i^T \right]
\end{eqnarray*}
Combining these expressions, the partial derivative of the information matrix can be expressed as a single (scaled) outer product:
\begin{eqnarray*}
\frac{\partial M(\boldsymbol{w})}{\partial w_i} &=& \frac{v_i}{\sigma_e^2} \left[ \boldsymbol{x}_i \boldsymbol{x}_i^T - \zeta_j \boldsymbol{x}_i\boldsymbol{S}_j^T - \zeta_j \boldsymbol{S}_j \boldsymbol{x}_i^T + \zeta_j^2 \boldsymbol{S}_j \boldsymbol{S}_j^T \right] \\ 
&=& \frac{v_i}{\sigma_e^2} \left(\boldsymbol{x}_i - \tfrac{\eta}{1 + \eta k_j} \boldsymbol{S}_j \right) \left( \boldsymbol{x}_i - \tfrac{\eta}{1 + \eta k_j} \boldsymbol{S}_j \right)^T
\end{eqnarray*}
The first-order Taylor approximation of the information matrix around $\boldsymbol{w}^{(0)}$ is:
\begin{eqnarray*}
\tilde{M}(\boldsymbol{w} \,|\, \boldsymbol{w}^{(0)}) & = & M(\boldsymbol{w}^{(0)}) + \sum_{i = 1}^n \big( w_i - w_i^{(0)} \big) \tfrac{\partial M(\boldsymbol{w})}{\partial w_i} \Big|_{\boldsymbol{w}^{(0)}} 
\end{eqnarray*}
Substituting expressions and reorganizing terms gives:
\begin{eqnarray*}
\sigma_e^2 \tilde{M}(\boldsymbol{w} \,|\, \boldsymbol{w}^{(0)}) & = & \sum_{j = 1}^b \sum_{i \in B_j} w_i v_i \left(\boldsymbol{x}_i - \zeta_j^{(0)}\boldsymbol{S}_j^{(0)}\right)\left(\boldsymbol{x}_i - \zeta_j^{(0)} \boldsymbol{S}_j^{(0)}\right)^T \\ 
&& - \sum_{j = 1}^b \zeta_j^{(0)} \boldsymbol{S}^{(0)}\left(\boldsymbol{S}_j^{(0)}\right)^T \\
&& + \sum_{j = 1}^b \zeta_j^{(0)}\left( \sum_{i \in B_j} w_i^{(0)} v_i \boldsymbol{x}_i \right) \left(\boldsymbol{S}_j^{(0)}\right)^T + \sum_{j = 1}^b \zeta_j^{(0)} \boldsymbol{S}_j^{(0)} \left( \sum_{i \in B_j} w_i^{(0)} v_i \boldsymbol{x}_i^T \right) \\
&& - \sum_{j=1}^b \left(\zeta_j^{(0)}\right)^2 \left( \sum_{i \in B_j} w_i^{(0)}v_i \right) \boldsymbol{S}_j^{(0)} \left(\boldsymbol{S}_j^{(0)}\right)^T
\end{eqnarray*}
Using that $k^{(0)}_j = \sum_{i \in B_j} w^{(0)}_iv_i$ and $\boldsymbol{S}_j^{(0)} = \sum_{i \in B_j} w^{(0)}_iv_i \boldsymbol{x}_i$, the above expression simplifies to:
\begin{eqnarray*}
\sigma_e^2 \tilde{M}(\boldsymbol{w} \,|\, \boldsymbol{w}^{(0)}) & = & \sum_{j = 1}^b \sum_{i \in B_j} w_i v_i \left(\boldsymbol{x}_i - \zeta_j^{(0)}\boldsymbol{S}_j^{(0)}\right)\left(\boldsymbol{x}_i - \zeta_j^{(0)} \boldsymbol{S}_j^{(0)}\right)^T \\ 
&& + \sum_{j = 1}^b \zeta_j^{(0)} \left( 1 - \zeta_j^{(0)} k_j^{(0)} \right) \boldsymbol{S}_j^{(0)} \left(\boldsymbol{S}_j^{(0)}\right)^T \\
&=&  \sum_{j = 1}^b \left[ \sum_{i \in B_j} w_i v_i \Big(\boldsymbol{x}_i - \tfrac{\eta}{1 + \eta k^{(0)}_j} \boldsymbol{S}^{(0)}_j \Big) \Big( \boldsymbol{x}_i - \tfrac{\eta}{1 + \eta k^{(0)}_j} \boldsymbol{S}^{(0)}_j \Big)^T + \tfrac{\eta}{(1 + \eta k_j^{(0)})^2} \boldsymbol{S}_j^{(0)} \Big(\boldsymbol{S}_j^{(0)}\Big)^T \right]
\end{eqnarray*}

\subsection{Optimal design augmentation}

\subsubsection{Approximate optimal design}

Given the initial approximate design $\boldsymbol{w}_0$, the approximate optimal augmented design solves:
\begin{eqnarray} \label{app_eq:2}
\boldsymbol{w}^* \ = \ \arg \max_{\boldsymbol{w} \in \Delta^n} \Phi(M(\boldsymbol{w}\,|\, \boldsymbol{w}_0) + \lambda I) - \alpha \cdot \boldsymbol{c}^T\boldsymbol{w}
\end{eqnarray}
where, for a linear model, the conditional information matrix (conditional on the error variance) is formulated as:
\begin{eqnarray}
M(\boldsymbol{w}\,|\, \boldsymbol{w}_0) \ = \ (1 - \gamma) M(\boldsymbol{w}_0 \,|\, \boldsymbol{X}, \boldsymbol{v}) + \gamma M(\boldsymbol{w} \,|\, \boldsymbol{X}, \boldsymbol{v}) \ = \ \frac{1}{\sigma_e^2} \sum_{i=1}^n \tilde{w}_i v_i \boldsymbol{x}_i \boldsymbol{x}_i^T,
\end{eqnarray}
with $\tilde{w}_i = (1-\gamma) w_{i, 0} + \gamma w_i$. The augmented optimal design problem in eq.(\ref{app_eq:2}) can be solved using the same SOCP formulation as for the non-augmented optimal design problem in Section \ref{sec:2.1} by imposing the following (convex) weight lower bound constraints to ensure that the optimized design weights $w^*_i$ remain nonnegative:
\begin{eqnarray}
\tilde{w}_i \geq (1 - \gamma) w_{i, 0}, \quad \text{for } i = 1,\ldots,n
\end{eqnarray}
where both $\gamma$ and $\boldsymbol{w}_0$ are given.

From the weights $\tilde{\boldsymbol{w}}^*$ returned by the SOCP solver, the weights $\boldsymbol{w}^*$ that optimize eq.(\ref{app_eq:2}) are obtained as:
\begin{eqnarray}
w^*_i \ = \ \frac{1}{\gamma} \tilde{w}^*_i - \frac{1 - \gamma}{\gamma} w_{i, 0} \ \geq \ 0, \quad \text{for } i = 1,\ldots, n
\end{eqnarray}
No extra constraint is needed to guarantee that $\sum_{i=1}^n w_i = 1$, since:
\begin{eqnarray}
\sum_{i = 1}^n w_i \ = \ \frac{1}{\gamma} \sum_{i=1}^n \big[ (1-\gamma)w_{i, 0} + \gamma w_i \big] - \frac{1-\gamma}{\gamma} \sum_{i = 1}^n w_{i, 0} \ = \ 1,
\end{eqnarray} 
using that $\sum_i \tilde{w}_i = 1$ is a constraint in the SOCP and $\sum_i w_{i, 0} = 1$ by assumption. For the other model types discussed in Section \ref{sec:3}, the SOCP formulation of the non-augmented optimal design problem can be extended in the same way to solve the augmented optimal design problem. 

\subsubsection{Cost-aware general equivalence theorem}

The Lagrangian of the penalized design optimization problem in Section \ref{sec:2.1} subject to D-optimality with $\Phi(M) = \log(\det(M)^{1/p})$ can be formulated as:
\begin{eqnarray}
\mathcal{L}(\boldsymbol{w}, \nu, \boldsymbol{\tau}) = \frac{1}{p}\log(\det(M(\boldsymbol{w})) - \alpha \sum_{i} c_i w_i - \nu\Big(\sum_i w_i - 1 \Big) + \sum_i \tau_i w_i,
\end{eqnarray}
using a multiplier $\nu$ for the equality constraint $\sum_i w_i = 1$, and multipliers $\tau_i \geq 0$ for the inequality constraints $w_i \geq 0$. Using Jacobi's formula and the cyclic property of the trace, the partial derivatives of the log-determinant with respect to the weights $w_i$ are:
\begin{eqnarray}
\frac{\partial \log(\det(M(\boldsymbol{w}))}{\partial w_i} & = & \text{Tr}\left[ M(\boldsymbol{w})^{-1} \frac{\partial M(\boldsymbol{w})}{\partial w_i} \right] \\
&=& \text{Tr}\left[ M(\boldsymbol{w})^{-1} \left( \frac{1}{\sigma_e^2} v_i \boldsymbol{x}_i \boldsymbol{x}_i^T \right) \right] \\
&=& \frac{1}{\sigma_e^2} v_i \boldsymbol{x}_i^T M(\boldsymbol{w})^{-1} \boldsymbol{x}_i,
\end{eqnarray}
where the term $\tfrac{1}{\sigma_e^2} \boldsymbol{x}_i^T M(\boldsymbol{w})^{-1} \boldsymbol{x}_i$ is exactly the standardized prediction variance $\sigma^2(\boldsymbol{x}_i\,|\,\boldsymbol{w})$ at the point $\boldsymbol{x}_i$. At the optimal design $\boldsymbol{w}^*$, by the Karush-Kuhn-Tucker (KKT) stationary condition, it must hold that $\frac{\partial\mathcal{L}}{\partial w_i} = 0$ for each $i$. That is,
\begin{eqnarray}
\frac{\partial \mathcal{L}}{\partial w_i} \ = \ \frac{1}{p} v_i \sigma^2(\boldsymbol{x}_i\,|\, \boldsymbol{w}^*) - \alpha c_i - \nu + \tau_i = 0, \quad \forall i
\end{eqnarray}
Furthermore, by the (KKT) complementary slackness condition, $\tau_i w_i = 0$ for each $i$, which leads to the following general equivalence theorem equations:
\begin{eqnarray}
\begin{cases}
\frac{1}{p} v_i \sigma^2(\boldsymbol{x}_i\,|\,\boldsymbol{w}^*) - \alpha c_i \ = \ \nu, & \text{if } w^*_i > 0,  \\
\frac{1}{p} v_i \sigma^2(\boldsymbol{x}_i\,|\,\boldsymbol{w}^*) - \alpha c_i \ = \ \nu - \tau_i \leq \nu, & \text{if } w^*_i = 0.
\end{cases}
\end{eqnarray}

\subsubsection{Gaussian process models}

Both active learning and Bayesian optimization typically use a Gaussian process as a surrogate model for the response of interest. The standard Gaussian process regression model with response vector $\boldsymbol{y} \in \mathbb{R}^{m \times 1}$ can be approximated by a Bayesian linear model \citep[Chapter 2]{RW06} of the form:
\begin{eqnarray}
\boldsymbol{y} \ \sim \ \Phi(\boldsymbol{X}, \boldsymbol{Z}) \boldsymbol{\theta} + \boldsymbol{\epsilon}, \quad \boldsymbol{\epsilon} \sim N(\boldsymbol{0}, \sigma_e^2 I_m)
\end{eqnarray}
with the vector $\boldsymbol{\theta} \sim N(\boldsymbol{0}, I_p)$ having a standard normal prior. The matrix $\Phi(\boldsymbol{X}, \boldsymbol{Z}) \in \mathbb{R}^{m \times p}$ is the projection of the feature matrix $\boldsymbol{X} \in \mathbb{R}^{m \times d}$ onto the span of a finite set of inducing points $\boldsymbol{Z} \in \mathbb{R}^{p \times d}$ using the Gaussian process kernel $k(\boldsymbol{x}, \boldsymbol{z})$. Specifically, $\Phi(\boldsymbol{X},\boldsymbol{Z})$ consists of row vectors $\phi(x)^T$, with:
\begin{eqnarray} 
\phi_Z(\boldsymbol{x}) \ = \ K_{ZZ}^{-1/2} k(\boldsymbol{Z}, \boldsymbol{x})
\end{eqnarray}
Here, $K_{ZZ}^{-1/2}$ is an inverse square root matrix of the kernel matrix $K_{ZZ} \in \mathbb{R}^{p \times p}$ with elements $K_{ij} = k(\boldsymbol{z}_i, \boldsymbol{z}_j)$; and $k(\boldsymbol{Z}, \boldsymbol{x}) \in \mathbb{R}^{p \times 1}$ is the cross-kernel vector between $\boldsymbol{Z}$ and $\boldsymbol{x}$ with elements $k_i = k(\boldsymbol{z}_i, \boldsymbol{x})$. If $\boldsymbol{Z} = \boldsymbol{X}$, $\Phi(\boldsymbol{X}, \boldsymbol{X})$ simplifies to a square root of the kernel matrix $K_{XX} \in \mathbb{R}^{m \times m}$ and the posterior mean of the Bayesian linear model coincides with that of the GPR model. However, the predictive variance at a new point $\boldsymbol{x}^*$ is $k(\boldsymbol{x}^*, \boldsymbol{X}) K_{XX}^{-1} k(\boldsymbol{X}, \boldsymbol{x}^*)$ which tends to zero when extrapolating away from the span of $\boldsymbol{X}$, whereas the GPR predictive variance tends to $k(\boldsymbol{x}^*, \boldsymbol{x}^*)$. Therefore, to exactly align the two models, draws from the GP posterior at $\boldsymbol{x}^*$ should be sampled as:
\begin{eqnarray}
f(\boldsymbol{x}^*) \ \sim \ \phi_X(\boldsymbol{x}^*)^T \boldsymbol{\theta} + \delta(\boldsymbol{x}^*), \quad \delta(\boldsymbol{x}^*) \sim N\left(0, k(\boldsymbol{x}^*, \boldsymbol{x}^*) - k(\boldsymbol{x}^*, \boldsymbol{X}) K_{XX}^{-1} k(\boldsymbol{X}, \boldsymbol{x}^*) \right).
\end{eqnarray}
Suppose that $\boldsymbol{\xi}$ is an exact design of size $m$ producing the feature matrix $\boldsymbol{X} \in \mathbb{R}^{m \times d}$, by eq.(\ref{eq:23}) the posterior information matrix $M(\boldsymbol{\xi})$ of the (approximate) Gaussian process regression model is given by:
\begin{eqnarray}
M(\boldsymbol{\xi}) \ = \ \frac{1}{\sigma_e^2} \Phi(\boldsymbol{X}, \boldsymbol{Z})^T \Phi(\boldsymbol{X}, \boldsymbol{Z}) + I_p
\end{eqnarray}
This means that penalized design augmentation can be performed for GPR models in the same way as for linear models using the feature map $\Phi(\boldsymbol{X}, \boldsymbol{Z})$ as the linear model design matrix. An advantage of GPR design augmentation in a linear model context is that the augmented design can be optimized jointly over the candidate design grid, whereas multi-dimensional discrete optimization of the acquisition function in active learning and Bayesian optimization is typically avoided due to the exploding computational effort. Note that for regression models with noise, i.e. $\sigma_e^2 > 0$, a ridge penalty with $\lambda = \sigma_e^2$ should be added to the diagonal of the linear model information matrix when optimizing the design to match the above formulation for $M(\boldsymbol{\xi})$.

\section{OMARS designs} \label{app:B}

Tables \ref{app_tab:1} to \ref{app_tab:3} list the 5-factor, 6-factor and 7-factor OMARS designs with correlation maps as shown in Figure \ref{fig:4}. The designs were obtained using the following code template (for a 5-factor design) adapted as needed for 6- and 7-factor designs:
\begin{knitrout}
\definecolor{shadecolor}{rgb}{1, 1, 1}\color{fgcolor}\begin{kframe}
\begin{alltt}
\hldef{> }\hldef{candidates} \hlkwb{<-} \hlkwd{expand.grid}\hldef{(}
\hldef{+ }  \hlkwc{x1} \hldef{=} \hlopt{-}\hlnum{1}\hlopt{:}\hlnum{1}\hldef{,}
\hldef{+ }  \hlkwc{x2} \hldef{=} \hlopt{-}\hlnum{1}\hlopt{:}\hlnum{1}\hldef{,}
\hldef{+ }  \hlkwc{x3} \hldef{=} \hlopt{-}\hlnum{1}\hlopt{:}\hlnum{1}\hldef{,}
\hldef{+ }  \hlkwc{x4} \hldef{=} \hlopt{-}\hlnum{1}\hlopt{:}\hlnum{1}\hldef{,}
\hldef{+ }  \hlkwc{x5} \hldef{=} \hlopt{-}\hlnum{1}\hlopt{:}\hlnum{1}
\hldef{+ }\hldef{)}
\hldef{> }\hldef{design} \hlkwb{<-} \hldef{lm_design}\hlopt{$}\hlkwd{new}\hldef{(}
\hldef{+ }  \hlkwc{formula} \hldef{=} \hlopt{~} \hlnum{0} \hlopt{+} \hldef{(x1} \hlopt{+} \hldef{x2} \hlopt{+} \hldef{x3} \hlopt{+} \hldef{x4} \hlopt{+} \hldef{x5)}\hlopt{^}\hlnum{2} \hlopt{+}
\hldef{+ }    \hlkwd{I}\hldef{(x1}\hlopt{^}\hlnum{2}\hldef{)} \hlopt{+} \hlkwd{I}\hldef{(x2}\hlopt{^}\hlnum{2}\hldef{)} \hlopt{+} \hlkwd{I}\hldef{(x3}\hlopt{^}\hlnum{2}\hldef{)} \hlopt{+} \hlkwd{I}\hldef{(x4}\hlopt{^}\hlnum{2}\hldef{)} \hlopt{+} \hlkwd{I}\hldef{(x5}\hlopt{^}\hlnum{2}\hldef{),}
\hldef{+ }  \hlkwc{orthogonal} \hldef{= x1} \hlopt{+} \hldef{x2} \hlopt{+} \hldef{x3} \hlopt{+} \hldef{x4} \hlopt{+} \hldef{x5} \hlopt{~} \hldef{.} \hlopt{~} \hlnum{0}\hldef{,}
\hldef{+ }  \hlkwc{zeros} \hldef{=} \hlkwd{list}\hldef{(}
\hldef{+ }    \hlopt{~} \hldef{(x1} \hlopt{+} \hldef{x2} \hlopt{+} \hldef{x3} \hlopt{+} \hldef{x4} \hlopt{+} \hldef{x5)}\hlopt{^}\hlnum{2}\hldef{,}
\hldef{+ }    \hlopt{~} \hldef{x1} \hlopt{+} \hldef{x2} \hlopt{+} \hldef{x3} \hlopt{+} \hldef{x4} \hlopt{+} \hldef{x5} \hlopt{+} \hlkwd{I}\hldef{(x1}\hlopt{^}\hlnum{2}\hldef{)} \hlopt{+} \hlkwd{I}\hldef{(x2}\hlopt{^}\hlnum{2}\hldef{)} \hlopt{+} \hlkwd{I}\hldef{(x3}\hlopt{^}\hlnum{2}\hldef{)} \hlopt{+} \hlkwd{I}\hldef{(x4}\hlopt{^}\hlnum{2}\hldef{)} \hlopt{+} \hlkwd{I}\hldef{(x5}\hlopt{^}\hlnum{2}\hldef{)}
\hldef{+ }  \hldef{),}
\hldef{+ }  \hlkwc{col_means} \hldef{= x1} \hlopt{+} \hldef{x2} \hlopt{+} \hldef{x3} \hlopt{+} \hldef{x4} \hlopt{+} \hldef{x5} \hlopt{~} \hlnum{0}\hldef{,}
\hldef{+ }  \hlkwc{data} \hldef{= candidates}
\hldef{+ }\hldef{)}
\hldef{> }\hldef{design}\hlopt{$}\hlkwd{round}\hldef{(}
\hldef{+ }  \hlkwc{m} \hldef{=} \hlnum{12}\hldef{,}
\hldef{+ }  \hlkwc{method} \hldef{=} \hlsng{"orthogonal"}\hldef{,}
\hldef{+ }  \hlkwc{highs_control} \hldef{=} \hlkwd{list}\hldef{(}\hlkwc{time_limit} \hldef{=} \hlnum{1800}\hldef{,} \hlkwc{mip_max_improving_sols} \hldef{=} \hlnum{1L}\hldef{)}
\hldef{+ }\hldef{)}
\end{alltt}
\end{kframe}
\end{knitrout}
The formulas passed to the \code{zeros} argument assert that: (1) the number of zero entries in all two-way interaction columns is the same; (2) the number of zero entries in all main effects columns is the same. The number of zeros in these two assertions are denoted by $n_0^{IE}$ and $n_0^{ME}$ in \cite{NG20}. The formula passed to the \code{col_means} argument asserts that each of the main effects columns \code{x1} to \code{x5} average to zero under the exact design.

\begin{figure}[t]
\center
\includegraphics[width=0.9\linewidth]{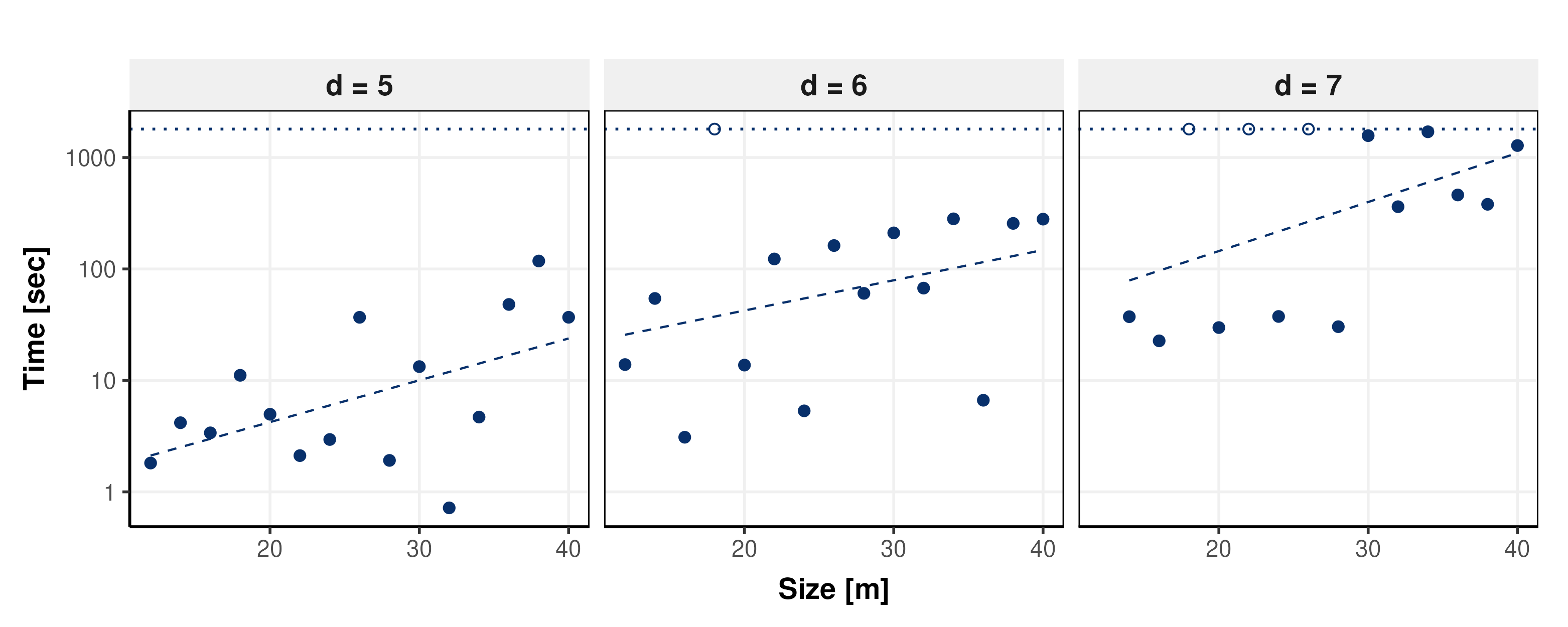}
\caption{Benchmark timings to round approximate to exact $d$-factor OMARS designs.} \label{app_fig:1}
\end{figure}

Figure \ref{app_fig:1} displays required computation times (in seconds) to generate exact OMARS designs of different sizes $m$. From left to right, the model corresponds to a $d$-factor full quadratic model, with $2d + d(d-1)/2$ design columns and candidate design grid of size $3^d$. The figure displays the computation times for the HiGHS solver (version 1.13.1) to find a single feasible solution, timed on a modern laptop computer (single-core Intel i7-8550U CPU). The maximum runtime was set to 1800 seconds, the design configurations that did not find any feasible solution within this time limit are highlighted by open circles.  

\section{Examples and R-code} \label{app:C}

\subsection{Generalized linear model example}

\begin{figure}[t]
\center
\includegraphics[width=1\linewidth]{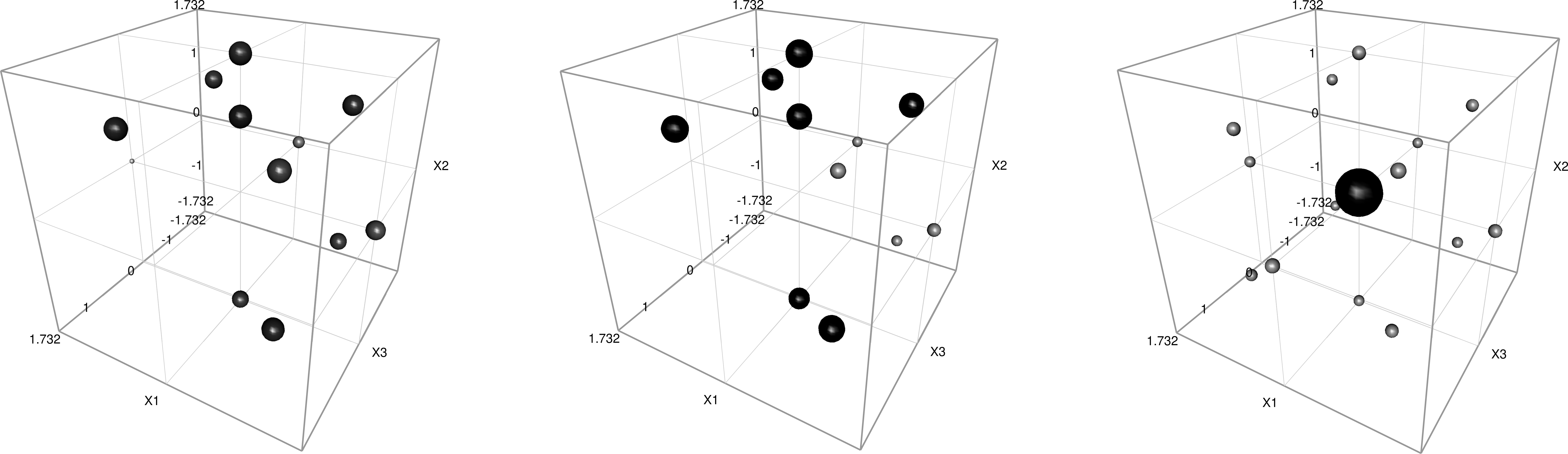}
\caption{Defect wafer experiment designs in the ($x_1$, $x_2$, $x_3$) space. Left: locally D-optimal approximate design. Middle: locally D-optimal exact design of size 18. Right: Spherical central composite design of size 18 (with 4 center points).} \label{app_fig:2}
\end{figure}

In \cite[Example 5]{LMM01}, a designed experiment is considered to count the number of defects in the manufacturing of semiconductor wafers. The defect counts are modeled by a generalized linear model with a Poisson distribution and log-link function. The systematic component consists of a second-order model in terms of three discrete factors $x_1$, $x_2$ and $x_3$:
\begin{eqnarray*}
\mu(\boldsymbol{x}) & = & g^{-1}(\eta(\boldsymbol{x})) \\
& = & \exp\big(\theta_0 + \theta_1 x_1 + \theta_2 x_2 + \theta_3 x_3 + \theta_4 x_1^2 + \theta_5 x_2^2 + \theta_6 x_3^2 + \theta_7 x_1 x_3 + \theta_8 x_1 x_3 + \theta_9 x_2 x_3 \big)
\end{eqnarray*}
An initial 18-run spherical central composite design is reported in \cite{LMM01}, where the factors are evaluated across 5 scaled levels: $\{ -1.732, -1, 0, 1, 1.732 \}$, which results in the parameter estimates $\boldsymbol{\theta}_0 = (1.9293, 0.6103, 0.7028, -0.3531, 0, 0.2471, 0, -0.5209, 0, 0)$. Given the model formula, nominal parameter values $\boldsymbol{\theta}_0$, and the candidate design grid, a locally D-optimal design of size $m = 18$ can be generated using the \proglang{R}-code below. Similar to the central composite design, the candidate design grid is constrained so that the extremes $\{ -1.732, 1.732 \}$ can only occur in combination with zeros for the remaining factors. In Figure \ref{app_fig:2}, this means that for each face of the cube, only the center of the face is included as a candidate design point. 
\begin{knitrout}
\definecolor{shadecolor}{rgb}{1, 1, 1}\color{fgcolor}\begin{kframe}
\begin{alltt}
\hldef{> }\hldef{candidates} \hlkwb{<-} \hlkwd{rbind}\hldef{(}
\hldef{+ }  \hlkwd{expand.grid}\hldef{(}\hlkwc{x1} \hldef{=} \hlopt{-}\hlnum{1}\hlopt{:}\hlnum{1}\hldef{,} \hlkwc{x2} \hldef{=} \hlopt{-}\hlnum{1}\hlopt{:}\hlnum{1}\hldef{,} \hlkwc{x3} \hldef{=} \hlopt{-}\hlnum{1}\hlopt{:}\hlnum{1}\hldef{),}
\hldef{+ }  \hlkwd{data.frame}\hldef{(}
\hldef{+ }    \hlkwc{x1} \hldef{=} \hlkwd{c}\hldef{(}\hlopt{-}\hlnum{1.732}\hldef{,} \hlnum{1.732}\hldef{,} \hlnum{0}\hldef{,} \hlnum{0}\hldef{,} \hlnum{0}\hldef{,} \hlnum{0}\hldef{),}
\hldef{+ }    \hlkwc{x2} \hldef{=} \hlkwd{c}\hldef{(}\hlnum{0}\hldef{,} \hlnum{0}\hldef{,} \hlopt{-}\hlnum{1.732}\hldef{,} \hlnum{1.732}\hldef{,} \hlnum{0}\hldef{,} \hlnum{0}\hldef{),}
\hldef{+ }    \hlkwc{x3} \hldef{=} \hlkwd{c}\hldef{(}\hlnum{0}\hldef{,} \hlnum{0}\hldef{,} \hlnum{0}\hldef{,} \hlnum{0}\hldef{,} \hlopt{-}\hlnum{1.732}\hldef{,} \hlnum{1.732}\hldef{)}
\hldef{+ }  \hldef{)}
\hldef{+ }\hldef{)}
\hldef{> }\hldef{design} \hlkwb{<-} \hldef{glm_design}\hlopt{$}\hlkwd{new}\hldef{(}
\hldef{+ }  \hlkwc{formula} \hldef{=} \hlopt{~} \hldef{(x1} \hlopt{+} \hldef{x2} \hlopt{+} \hldef{x3)}\hlopt{^}\hlnum{2} \hlopt{+} \hlkwd{I}\hldef{(x1}\hlopt{^}\hlnum{2}\hldef{)} \hlopt{+} \hlkwd{I}\hldef{(x2}\hlopt{^}\hlnum{2}\hldef{)} \hlopt{+} \hlkwd{I}\hldef{(x3}\hlopt{^}\hlnum{2}\hldef{),}
\hldef{+ }  \hlkwc{data} \hldef{= candidates,}
\hldef{+ }  \hlkwc{family} \hldef{=} \hlkwd{poisson}\hldef{(}\hlkwc{link} \hldef{=} \hlsng{"log"}\hldef{),}
\hldef{+ }  \hlkwc{theta} \hldef{=} \hlkwd{c}\hldef{(}\hlnum{1.9293}\hldef{,} \hlnum{0.6103}\hldef{,} \hlnum{0.7028}\hldef{,} \hlopt{-}\hlnum{0.3531}\hldef{,} \hlnum{0}\hldef{,} \hlnum{0.2471}\hldef{,} \hlnum{0}\hldef{,} \hlopt{-}\hlnum{0.5209}\hldef{,} \hlnum{0}\hldef{,} \hlnum{0}\hldef{)}
\hldef{+ }\hldef{)}
\hldef{> }\hldef{design}\hlopt{$}\hlkwd{optimize}\hldef{(}\hlkwc{criterion} \hldef{=} \hlsng{"D"}\hldef{)}
\hldef{> }\hldef{design}\hlopt{$}\hlkwd{round}\hldef{(}\hlkwc{m} \hldef{=} \hlnum{18}\hldef{,} \hlkwc{method} \hldef{=} \hlsng{"optimal"}\hldef{)}
\end{alltt}
\end{kframe}
\end{knitrout}
The exact D-optimal design is 15\% more efficient than the central composite design in terms of the ratio of D-criteria. This is not surprising, as the D-optimal design was generated conditional on the nominal parameter values $\boldsymbol{\theta}_0$. For comparison, a D-optimal design based on the systematic component viewed as a linear model is only 4.3\% more efficient than the central composite design.

\subsection{Optimal design augmentation}

\subsubsection{Imidazoles sequential design augmentation}

\begin{knitrout}
\definecolor{shadecolor}{rgb}{1, 1, 1}\color{fgcolor}\begin{kframe}
\begin{alltt}
\hldef{> }\hldef{candidates} \hlkwb{<-} \hlkwd{transform}\hldef{(}
\hldef{+ }  \hldef{imidazoles[,} \hlkwd{c}\hldef{(}\hlsng{"concentration"}\hldef{,} \hlsng{"temperature"}\hldef{,} \hlsng{"base"}\hldef{,} \hlsng{"solvent"}\hldef{,} \hlsng{"ligand"}\hldef{)],}
\hldef{+ }  \hlkwc{concentration} \hldef{= (concentration} \hlopt{-} \hlnum{0.057}\hldef{)} \hlopt{/} \hlnum{0.096}\hldef{,}
\hldef{+ }  \hlkwc{temperature} \hldef{= (temperature} \hlopt{-} \hlnum{90}\hldef{)} \hlopt{/} \hlnum{30}
\hldef{+ }\hldef{)}
\hldef{> }\hldef{design} \hlkwb{<-} \hldef{lm_design}\hlopt{$}\hlkwd{new}\hldef{(}
\hldef{+ }  \hlkwc{formula} \hldef{=} \hlopt{~} \hldef{(concentration} \hlopt{+} \hldef{temperature} \hlopt{+} \hldef{base} \hlopt{+} \hldef{solvent} \hlopt{+} \hldef{ligand)}\hlopt{^}\hlnum{2}\hldef{,}
\hldef{+ }  \hlkwc{data} \hldef{= candidates,}
\hldef{+ }  \hlkwc{lambda} \hldef{=} \hlnum{0.01}
\hldef{+ }\hldef{)}
\hldef{> }\hldef{current_design} \hlkwb{<-} \hldef{design}\hlopt{$}\hlkwd{round}\hldef{(}
\hldef{+ }  \hlkwc{m} \hldef{=} \hlnum{15}\hldef{,}
\hldef{+ }  \hlkwc{method} \hldef{=} \hlsng{"optimal"}\hldef{,}
\hldef{+ }  \hlkwc{n_repeats} \hldef{=} \hlnum{25}\hldef{,}
\hldef{+ }  \hlkwc{criterion} \hldef{=} \hlsng{"I"}\hldef{,}
\hldef{+ }  \hlkwc{design_weights} \hldef{=} \hlnum{FALSE}
\hldef{+ }\hldef{)}
\hldef{> }\hlkwa{for} \hldef{(iter} \hlkwa{in} \hlnum{1}\hlopt{:}\hlnum{6}\hldef{) \{}
\hldef{+ }  \hldef{yhat} \hlkwb{<-} \hlkwd{predict_placeholder}\hldef{(current_design)} \hlcom{## insert lasso predictions}
\hldef{+ }  \hldef{design}\hlopt{$}\hlkwd{update}\hldef{(}
\hldef{+ }    \hlkwc{lambda} \hldef{=} \hlnum{0.9} \hlopt{*} \hldef{design}\hlopt{$}\hldef{lambda,}
\hldef{+ }    \hlkwc{weights} \hldef{=} \hlkwd{pmax}\hldef{(yhat,} \hlnum{0}\hldef{)}
\hldef{+ }  \hldef{)}
\hldef{+ }  \hldef{current_design} \hlkwb{<-} \hldef{design}\hlopt{$}\hlkwd{round}\hldef{(}
\hldef{+ }    \hlkwc{m} \hldef{=} \hlnum{10}\hldef{,}
\hldef{+ }    \hlkwc{method} \hldef{=} \hlsng{"optimal"}\hldef{,}
\hldef{+ }    \hlkwc{augment_design} \hldef{= current_design,}
\hldef{+ }    \hlkwc{n_repeats} \hldef{=} \hlnum{25}\hldef{,}
\hldef{+ }    \hlkwc{criterion} \hldef{=} \hlsng{"I"}\hldef{,}
\hldef{+ }    \hlkwc{replicates} \hldef{=} \hlnum{FALSE}\hldef{,}
\hldef{+ }    \hlkwc{design_weights} \hldef{=} \hlnum{FALSE}
\hldef{+ }  \hldef{)}
\hldef{+ }\hldef{\}}
\end{alltt}
\end{kframe}
\end{knitrout}

\newpage

\section{Design tables} \label{app:D}

\subsection{OMARS designs}

\begin{table}[H]
\centering
\begin{tabular}{rrrrr}
\toprule
\code{x1} & \code{x2} & \code{x3} & \code{x4} & \code{x5}\\
\midrule
-1 & -1 & 0 & -1 & -1\\
1 & 1 & 1 & -1 & -1\\
1 & -1 & -1 & 0 & -1\\
0 & 1 & -1 & 1 & -1\\
-1 & 0 & 1 & 1 & -1\\
\addlinespace
-1 & 1 & -1 & -1 & 0\\
1 & -1 & 1 & 1 & 0\\
1 & 0 & -1 & -1 & 1\\
0 & -1 & 1 & -1 & 1\\
-1 & 1 & 1 & 0 & 1\\
\addlinespace
-1 & -1 & -1 & 1 & 1\\
1 & 1 & 0 & 1 & 1\\
\bottomrule
\end{tabular}
\caption{5-factor OMARS design of size $m=12$.} \label{app_tab:1}
\end{table}

\begin{table}[H]
\centering
\begin{tabular}{rrrrrr}
\toprule
\code{x1} & \code{x2} & \code{x3} & \code{x4} & \code{x5} & \code{x6}\\
\midrule
1 & 1 & 1 & -1 & -1 & -1\\
-1 & -1 & 1 & 0 & -1 & -1\\
1 & -1 & -1 & 1 & -1 & -1\\
-1 & 1 & -1 & -1 & 0 & -1\\
0 & -1 & -1 & -1 & 1 & -1\\
\addlinespace
1 & 1 & 0 & 1 & 1 & -1\\
-1 & 0 & 1 & 1 & 1 & -1\\
-1 & 1 & -1 & 1 & -1 & 0\\
1 & -1 & 1 & -1 & 1 & 0\\
1 & 0 & -1 & -1 & -1 & 1\\
\addlinespace
-1 & -1 & 0 & -1 & -1 & 1\\
0 & 1 & 1 & 1 & -1 & 1\\
1 & -1 & 1 & 1 & 0 & 1\\
-1 & 1 & 1 & -1 & 1 & 1\\
1 & 1 & -1 & 0 & 1 & 1\\
\addlinespace
-1 & -1 & -1 & 1 & 1 & 1\\
\bottomrule
\end{tabular}
\caption{6-factor OMARS design of size $m=16$.} \label{app_tab:2}
\end{table}

\begin{table}[H]
\centering
\begin{tabular}{rrrrrrr}
\toprule
\code{x1} & \code{x2} & \code{x3} & \code{x4} & \code{x5} & \code{x6} & \code{x7}\\
\midrule
-1 & 0 & -1 & -1 & -1 & -1 & -1\\
1 & -1 & 1 & -1 & -1 & -1 & -1\\
0 & -1 & -1 & 1 & 1 & -1 & -1\\
-1 & 1 & 1 & 1 & 1 & -1 & -1\\
1 & 1 & -1 & -1 & 1 & 0 & -1\\
\addlinespace
1 & 1 & -1 & 1 & -1 & 1 & -1\\
-1 & -1 & 0 & 1 & -1 & 1 & -1\\
-1 & 1 & 1 & -1 & 0 & 1 & -1\\
1 & -1 & 1 & 0 & 1 & 1 & -1\\
1 & 1 & 1 & 1 & -1 & -1 & 0\\
\addlinespace
-1 & -1 & -1 & -1 & 1 & 1 & 0\\
-1 & 1 & -1 & 0 & -1 & -1 & 1\\
1 & -1 & -1 & 1 & 0 & -1 & 1\\
1 & 1 & 0 & -1 & 1 & -1 & 1\\
-1 & -1 & 1 & -1 & 1 & -1 & 1\\
\addlinespace
-1 & -1 & 1 & 1 & -1 & 0 & 1\\
1 & -1 & -1 & -1 & -1 & 1 & 1\\
0 & 1 & 1 & -1 & -1 & 1 & 1\\
-1 & 1 & -1 & 1 & 1 & 1 & 1\\
1 & 0 & 1 & 1 & 1 & 1 & 1\\
\bottomrule
\end{tabular}
\caption{7-factor OMARS design of size $m=20$.} \label{app_tab:3}
\end{table}

\subsection{Split-plot design example case study}

\begin{table}[H]
\centering
\begin{tabular}{rrrrr}
\toprule
week & A & B & C & D\\
\midrule
1 & -1 & -1 & -1 & 1\\
1 & -1 & -1 & 1 & 1\\
1 & -1 & 1 & 1 & -1\\
2 & -1 & -1 & -1 & -1\\
2 & -1 & 0 & 0 & 0\\
2 & -1 & 1 & -1 & 1\\
3 & -1 & -1 & 1 & -1\\
3 & -1 & 1 & -1 & -1\\
3 & -1 & 1 & 1 & 1\\
4 & 1 & -1 & 0 & -1\\
4 & 1 & 0 & -1 & 1\\
4 & 1 & 1 & 1 & 0\\
5 & 1 & -1 & 1 & 1\\
5 & 1 & 0 & 1 & -1\\
5 & 1 & 1 & -1 & -1\\
6 & 0 & -1 & 0 & 1\\
6 & 0 & 0 & -1 & -1\\
6 & 0 & 1 & -1 & 0\\
7 & 1 & -1 & -1 & 0\\
7 & 1 & 1 & 0 & 1\\
7 & 1 & 1 & 1 & -1\\
\bottomrule
\end{tabular}
\caption{D-optimal split-plot design for the case study in \cite{OM25}.} \label{app_tab:4}
\end{table}

%\begin{table}[H]
%\centering
%\begin{tabular}{rrrrr}
%\toprule
%week & A & B & C & D\\
%\midrule
%1 & -1 & -1 & 1 & -1\\
%1 & -1 & 1 & 0 & 1\\
%1 & -1 & 1 & 1 & 0\\
%2 & 0 & -1 & -1 & -1\\
%2 & 0 & 1 & -1 & 1\\
%2 & 0 & 1 & 1 & -1\\
%3 & 1 & 0 & -1 & 1\\
%3 & 1 & 0 & 1 & -1\\
%3 & 1 & 1 & 0 & 0\\
%4 & 1 & -1 & -1 & -1\\
%4 & 1 & 0 & 0 & 0\\
%4 & 1 & 1 & 1 & 1\\
%5 & -1 & -1 & -1 & 1\\
%5 & -1 & 0 & 1 & 1\\
%5 & -1 & 1 & -1 & -1\\
%6 & 1 & -1 & 0 & 1\\
%6 & 1 & -1 & 1 & 0\\
%6 & 1 & 1 & -1 & -1\\
%7 & 0 & -1 & 1 & 1\\
%7 & 0 & 0 & -1 & 0\\
%7 & 0 & 0 & 0 & -1\\
%\bottomrule
%\end{tabular}
%\caption{I-optimal split-plot design for the case study in \cite{OM25}.} \label{app_tab:5}
%\end{table}

\subsection{Augmented extraction experiment}

\begin{table}[H]
\centering
\begin{tabular}{lrrrrrr}
\toprule
  & \code{x1} & \code{x2} & \code{x3} & \code{x4} & \code{x5} & \code{x6}\\
\midrule
1 & 1 & 1 & 1 & -1 & -1 & -1\\
2 & -1 & -1 & -1 & 1 & -1 & -1\\
3 & -1 & 1 & 1 & 1 & -1 & -1\\
4 & 1 & -1 & -1 & -1 & 1 & -1\\
5 & -1 & 1 & -1 & -1 & 1 & -1\\
6 & 1 & -1 & 1 & 1 & 1 & -1\\
7 & 1 & 1 & -1 & -1 & -1 & 1\\
8 & -1 & -1 & 1 & -1 & -1 & 1\\
9 & 1 & -1 & -1 & 1 & -1 & 1\\
10 & -1 & -1 & 1 & -1 & 1 & 1\\
11 & -1 & 1 & -1 & 1 & 1 & 1\\
12 & 1 & 1 & 1 & 1 & 1 & 1\\
\midrule
13 & 1 & -1 & -1 & -1 & -1 & -1\\
14 & -1 & -1 & 1 & -1 & -1 & -1\\
15 & 1 & 1 & -1 & 1 & -1 & -1\\
16 & -1 & 1 & 1 & 1 & 1 & -1\\
17 & -1 & 1 & 1 & -1 & -1 & 1\\
18 & -1 & -1 & 1 & 1 & -1 & 1\\
19 & -1 & -1 & -1 & -1 & 1 & 1\\
20 & 1 & -1 & 1 & -1 & 1 & 1\\
\bottomrule
\end{tabular}
\caption{Augmented extraction experiment: initial design of size $m_0 = 12$ and design augmentation of size $m_1 = 8$ from \cite[Chapter 3]{GJ11}.} \label{app_tab:6}
\end{table}

\end{appendix}

\newpage

%% -----------------------------------------------------------------------------

\end{document}